\documentclass[apj]{emulateapj}

\newcommand{\arcm}{\ifmmode {' }\else $' $\fi}
\newcommand{\arcs}{\ifmmode {'' }\else $'' $\fi}

\slugcomment{}

\shortauthors{Rhode et al.} \shorttitle{Globular Cluster Systems of
  Four Spiral Galaxies}

\begin{document}

\title{Global Properties of the Globular Cluster Systems \\ of Four Spiral Galaxies}

\author{Katherine L. Rhode\altaffilmark{1}}
\affil{Astronomy Department, Wesleyan University, Middletown, CT
  06459; kathy@astro.wesleyan.edu\\
and\\
Department of Astronomy, Yale University, New
  Haven, CT 06520}

\author{Stephen E. Zepf}
\affil{Department of Physics \& Astronomy, Michigan State University,
  East Lansing, MI 48824; zepf@pa.msu.edu}

\author{Arunav Kundu}
\affil{Department of Physics \& Astronomy, Michigan State University,
  East Lansing, MI 48824; kundu@pa.msu.edu}

\author{Aaron N. Larner}
\affil{Astronomy Department, Wesleyan University, Middletown, CT
  06459; alarner@wesleyan.edu}

\altaffiltext{1}{NSF Astronomy \& Astrophysics Postdoctoral Fellow}

\begin{abstract}
We present results from a wide-field imaging study of the globular
cluster (GC) systems of a sample of edge-on, Sb$-$Sc spiral galaxies
$\sim$7-20~Mpc away.  This study is part of a larger survey of the
ensemble properties of the GC populations of giant galaxies.  We
imaged the galaxies in $BVR$ filters with large-format CCD detectors
on the WIYN 3.5-m telescope, to projected radii of $\sim$20$-$40~kpc.
For four galaxies (NGC~2683, NGC~3556, NGC~4157, and NGC~7331), we
quantify the radial distributions of the GC systems and estimate the
total number, luminosity- and mass-normalized specific frequencies
($S_N$ and $T$), and blue (metal-poor) fraction of GCs.  A fifth
galaxy (NGC~3044) was apparently too distant for us to have detected
its GC system.  Our $S_N$ for NGC~2683 is 2.5 times smaller than the
previously-published value, likely due in part to reduced
contamination from non-GCs.  For the spiral galaxies analyzed for the
survey to date, the average number of GCs is 170$\pm$40 and the
weighted mean values of $S_N$ and $T$ are 0.8$\pm$0.2 and 1.4$\pm$0.3.
We use the survey data to derive a relationship between radial exent
of the GC system and host galaxy mass over a factor of 20 in mass.
Finally, we confirm the trend, identified in previous survey papers,
of increasing specific frequency of metal-poor GCs with increasing
galaxy mass.  We compare the data with predictions from a simple model
and show that carefully quantifying the numbers of metal-poor GCs in
galaxies can constrain the formation redshifts of the GCs and their
host galaxies.
\end{abstract}


\section{Introduction}
\label{section:introduction}
Much has been learned about the structure and formation of the Milky
Way Galaxy from studies of its globular cluster system.  The key
historical development in this area was Shapley's use of globular
clusters to investigate the structure of the Galaxy and the location
of the Solar System within it (e.g., Shapley 1918).  Many decades
later, \citet{sz78} made abundance estimates for a subset of Galactic
globular clusters and used them to infer a chaotic, hierarchical
scenario for the formation of the Galactic halo.  \citet{zinn85} later
identified distinct ``halo'' and ``disk'' subpopulations of globular
clusters in the Milky Way, having different kinematics, spatial
distributions, and, by inference, different origins.  To date,
$\sim$150 globular clusters have been identified in our Galaxy
\citep{harris96} and numerous studies have yielded estimates of their
distances, abundances, kinematics, and ages; together these provide
crucial information regarding the assembly history of the Galaxy.
Furthermore, the globular cluster system of the other massive spiral
galaxy in our neighborhood, Andromeda, has been surveyed fairly
completely in the past decade, so that we now have estimates of the
colors, metallicities, and kinematics of a substantial fraction of its
$\sim$450 globular clusters (e.g., Barmby et al.\ 2000; Perrett et
al.\ 2002).

The natural question that arises is whether what we have learned about
the globular cluster systems of the Milky Way and Andromeda is true
for other galaxies of similar mass, especially spiral galaxies.  Are
the Milky Way and Andromeda representative of other spiral galaxies,
in terms of the total numbers, spatial distributions, colors, and
specific frequencies of their globular cluster populations?  If the
Milky Way and Andromeda globular cluster systems are similar to (or
different from) those of other galaxies, what does that tell us about
how galaxies form and evolve?

Although studies of extragalactic globular cluster systems have
multiplied rapidly over the past decade (see the recent review by
Brodie \& Strader 2006), observational studies of spiral galaxy GC
systems are still comparatively rare.  \citet{az98} put together a
comprehensive table of the existing data on galaxies' GC systems
(quantities such as total number, specific frequency, and mean
metallicity).  The table included 82 galaxies, and only twelve were
spiral galaxies (Hubble type Sa $-$ Scd), including the Milky Way and
M31.  Since that time, {\em Hubble Space Telescope (HST)} studies of
several more spiral galaxies have been published (e.g., Goudfrooij et
al.\ 2003; Chandar, Whitmore, \& Lee 2004).  Although the high
resolution of $HST$ offers distinct advantages in terms of
distinguishing GCs from contaminants such as faint background
galaxies, its small field of view means that typically only a small
subset of the area around the galaxies is observed, which makes it
difficult to accurately determine the global properties (spatial and
color distributions, total numbers) of the GC systems.  For example,
we showed in our wide-field imaging study of the GC system of the
spiral galaxy NGC~7814 that quantities like the total number and
specific frequency of GCs can be off by $\sim$20$-$75\% when one
extrapolates results from {\it HST} data, or small-format CCD data,
out to large radius (Rhode \& Zepf 2003 and references therein).

This paper presents results from wide-field CCD imaging of the
globular cluster systems of four Sb$-$Sc spiral galaxies: NGC~2683,
NGC~3556, NGC~4157, and NGC~7331.  We also discuss observations of a
fifth galaxy, the Sc galaxy NGC~3044, which apparently is too distant
for us to have detected its GC system.  Basic properties of these five
galaxies are given in Table~\ref{table:galaxy properties}. The data
presented here were acquired as part of a survey that uses
large-format and mosaic CCD imagers to study the global properties of
the GC systems of spiral, S0, and elliptical galaxies at distances of
$\sim$7$-$20~Mpc.  A description of the survey, and results for the
first five galaxies analyzed, are given in Rhode \& Zepf (2001, 2003,
and 2004; hereafter RZ01, RZ03, RZ04).  Because of difficulties in
quantifying the selection effects caused by intrinsic structure and
line-of-sight extinction in spiral galaxy disks, the only reliable way
to quantify the global properties of spiral galaxy GC populations is
to study galaxies that appear edge-on in the sky.  Therefore the
spiral galaxy targets chosen for the survey have
$i$~$_<\atop{^\sim}$75$^\circ$.  We use techniques such as imaging in
multiple filters and analyzing archival $HST$ data to carefully reduce
contamination from non-GCs and estimate the amount of contamination
that remains in the samples.  Our main goal is to accurately quantify
the spatial distribution of each galaxy's GC system over its full
radial range, in order to then calculate a reliable total number of
GCs for the system.  We can then compare these total numbers to
predictions from galaxy formation models such as Ashman \& Zepf (1992;
hereafter AZ92), who suggested that elliptical galaxies and their GC
systems can be formed by the collision of spiral galaxies.  Somewhat
more generally, we wish to compare the global properties of the GC
systems of the spiral galaxies in the survey to those of galaxies of
other morphological types (ellipticals and S0s).  Making such a
comparison will help us determine the typical GC system properties for
galaxies of different types, and what that might tell us about galaxy
origins in general.

The paper is organized as follows.  Section~\ref{section:reductions}
describes the observations and initial data reduction steps.
Section~\ref{section:analysis} explains our methods for detecting GCs
and analyzing the GC system properties.  Section~\ref{section:results}
gives the results, and the final section is a summary of the study.

\section{Observations \& Initial Reductions}
\label{section:reductions}

Images of the targeted spiral galaxies were taken between 1999 October
  and 2001 January with the 3.5-m WIYN telescope\footnote{The WIYN
  Observatory is a joint facility of the University of Wisconsin,
  Indiana University, Yale University, and the National Optical
  Astronomy Observatories.} at Kitt Peak National Observatory.  One of
  two CCD detectors was used; either a single 2048 x 2048-pixel CCD
  (S2KB) with 0.196$\arcsec$ pixels and a 6.7$\arcmin$~x~6.7$\arcmin$
  field of view, or the Minimosaic Imager, which consists of two 2048
  x 4096-pixel CCD detectors with 0.14$\arcsec$ pixels and a total
  field of view 9.6$\arcmin$~x~9.6$\arcmin$.  For the four nearest
  galaxies (with distances 7$-$15~Mpc), the galaxy was positioned
  toward the edge of the detector, to maximize the radial coverage of
  the GC system.  NGC~3044 was substantially more distant, at
  $>$20~Mpc away, and so was positioned in the center of the detector.
  A series of images was taken in three broadband filters ($BVR$).
  Table~\ref{table:wiyn observations} specifies for each galaxy the
  dates of the observations, the detector used, and the number of
  exposures and integration times in each filter.

The images of NGC~3044, NGC~3556, NGC~4157, and NGC~7331 were taken
under photometric conditions, and calibrated with observations of
photometric standard stars \citep{land92} taken on the same nights as
the imaging data.  The images of NGC~2683 were taken under
non-photometric sky conditions.  In this case we took single, short
(400 $-$ 600~s) $BVR$ exposures of the galaxy on a subsequent,
photometric night during the same observing run (in January 2001).  We
used these short exposures along with calibration frames taken on the
same night to post-calibrate the longer exposures of NGC~2683.

The photometric calibration data for the five galaxies discussed in
this paper were taken on five different nights, with one of two CCD
detectors (Minimosaic or S2KB).  The color coefficients in the $V$
magnitude equation ranged from 0.02 to 0.08, with a typical formal
uncertainty of 0.01.  The color coefficients in the $B-V$ color
equation ranged from 1.01 to 1.06, with a typical uncertainty of 0.01.
The color coefficients in the $V-R$ color equation ranged from 1.04 to
1.06, with an uncertainty of 0.01 to 0.03.  The formal errors on the
zero points in the $V$, $B-V$, and $V-R$ calibration equations fell
between 0.003 and 0.01, indicating that it was in fact photometric on
the nights that the calibration data were taken.

Preliminary reductions (overscan and bias subtraction, flat-field
division) were done with standard reduction tasks in the
IRAF\footnote{IRAF is distributed by the National Optical Astronomy
Observatories, which are operated by the Association of Universities
for Research in Astronomy (AURA), Inc., under cooperative agreement
with the National Science Foundation.} packages CCDRED (for the S2KB
images) or MSCRED (for the Minimosaic images).  The MSCRED tasks
MSCZERO, MSCCMATCH, and MSCIMAGE were used to convert the
multi-extension Minimosaic FITS images into single images. The images
taken of each galaxy target were aligned to each other, and then sky
subtraction was performed on each individual image.  The individual
images taken with a given filter of a given galaxy target were then
scaled to a common flux level and combined, to create a deep,
cosmic-ray-free image of each galaxy in each of the three filters.
Finally, the sky background level was restored to each of the combined
images.  The resolution (FWHM of the point spread function) of the
final combined images ranges from: 0.6$\arcsec$ to 0.9$\arcsec$ for
NGC~2683, 0.7$\arcsec$ to 1.1$\arcsec$ for NGC~3044; 0.7$\arcsec$ to
1.0$\arcsec$ for NGC~3556; 0.9$\arcsec$ to 1.1$\arcsec$ for NGC~4157;
and 0.9$\arcsec$ to 1.0$\arcsec$ for NGC~7814.

\section{Detection \& Analysis of the Globular Cluster System}
\label{section:analysis}

\subsection{Source Detection and Matching}
\label{section:source detection}
Globular clusters at the distances of our galaxy targets will appear
unresolved in ground-based images.  To detect GCs, we first removed
the diffuse galaxy light from the images.  The final combined images
were smoothed with a ring median filter of diameter equal to 7 times
the mean FWHM of point sources in the image.  The smoothed images were
then subtracted from the original versions to create a
galaxy-subtracted image.  (We experimented with filters of varying
diameter for the smoothing step and found that the ring filter with
the specified diameter consistently removed the diffuse galaxy light
without removing any of the light from the point sources.)  The
appropriate constant sky level was restored to the galaxy-subtracted
images and then the IRAF task DAOFIND was used to detect sources with
a signal-to-noise level between 3.5 and 6 times the noise in the
background.  We masked out the high-noise regions of the
galaxy-subtracted images where pointlike GC candidates could not
reliably be detected, such as the inner, dusty disks of the galaxies
and regions immediately surrounding saturated foreground stars or
large resolved background galaxies.  We removed from the DAOFIND lists
any sources located within these masked regions, and then matched the
remaining sources to produce a final list of sources detected in all
three filters.  The number of sources remaining after this step was
537 in NGC~2683, 522 in NGC~3044, 573 in NGC~3556, 387 in NGC~4157,
and 304 in NGC~7331.

\subsection{Eliminating Extended Sources}

\begin{figure}
\plotone{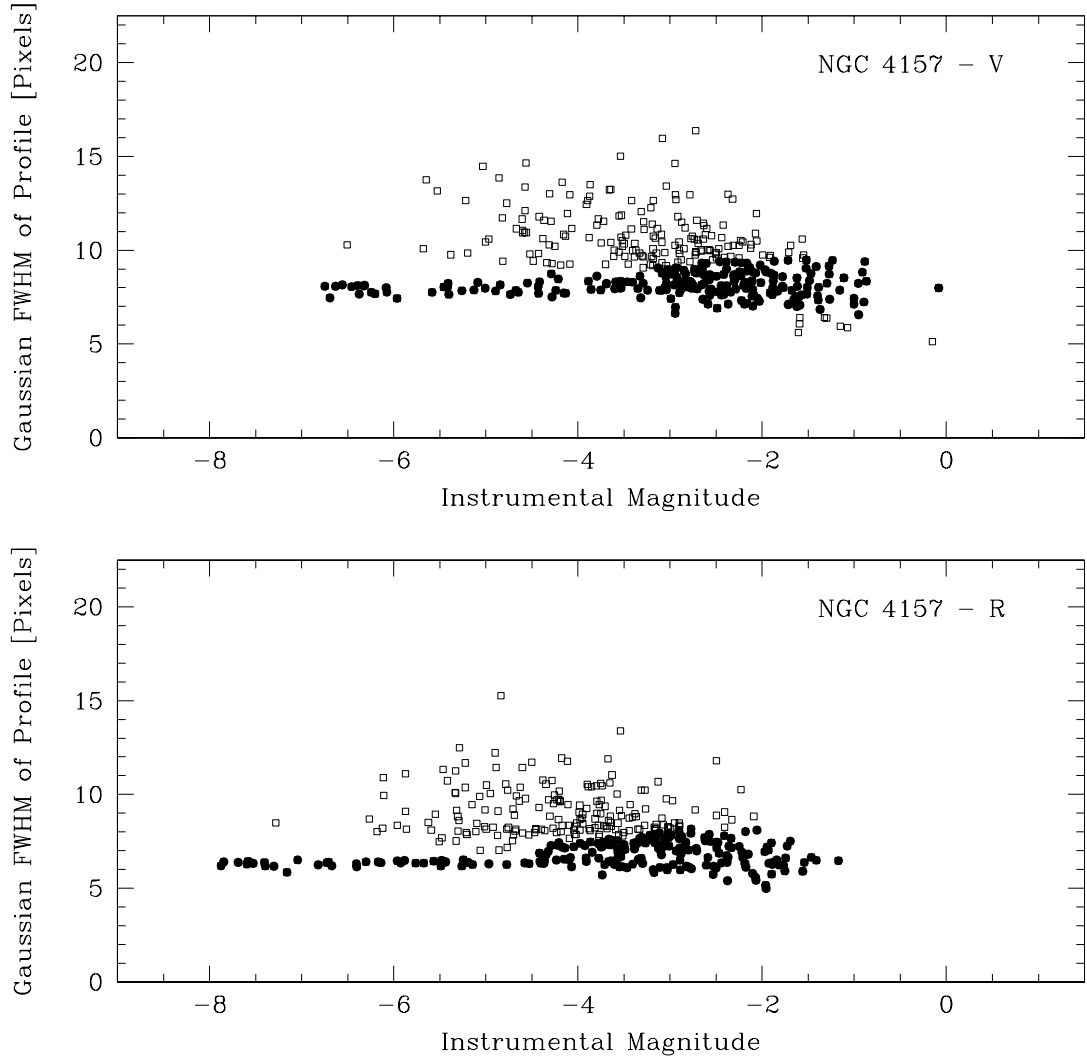}
\caption{\normalsize Gaussian FWHM of the radial profile vs.\ instrumental
  magnitude for 387 objects in the WIYN $V$ and $R$ images of
  NGC~4157.  Filled circles are objects that pass the FWHM criteria;
  open squares are those that are not likely to be point sources and
  are therefore excluded as GC candidates.}
\label{fig:fwhm mag}
\end{figure}

To remove extended objects (e.g., contaminating background galaxies)
from the source lists, we began by measuring the FWHM of each source
in the matched lists and plotting it versus its instrumental
magnitude.  An example of such a plot is Figure~\ref{fig:fwhm mag},
which shows FWHM vs. magnitude for the 387 detected sources in the $V$
and $R$ images of NGC~4157.  At bright magnitudes, point sources have
FWHM values that form a tight sequence around some mean value.
Extended objects have larger FWHM values, scattered over a larger
range.  At fainter instrumental magnitudes, point sources still
scatter around the same mean value, but their FWHM values spread out;
consequently, the border between the FWHM values of point sources and
extended objects becomes less clear at faint magnitudes.

We created FWHM vs. instrumental magnitude plots for the final
combined images of each galaxy, and then eliminated extended objects
by selecting as GC candidates only those sources with FWHM values
close to the mean FWHM value of point sources for each image.  We
visually examined the plots to determine the boundary between point
sources and extended objects, and then wrote a computer code to
implement the extended source cut.  The range of acceptable FWHM
values gradually increases with increasing magnitude of the sources.
We used the FWHM information in the different filters independently
(i.e., if a source had a large FWHM value in {\it one} of the combined
images, it was removed from the GC candidate lists).  For NGC~2683 and
NGC~7331, we used measurements in all three broadband filters to
determine whether a source was extended.  For the other three galaxies
(NGC~3044, NGC~3556, and NGC~4157), we used measurements from the $V$
and $R$ images only, since they had better resolution than the
$B$-band images and the FWHM versus magnitude plots showed
significantly less scatter in those filters compared to the plot made
from the $B$ image.

Figure~\ref{fig:fwhm mag} shows typical results from this source
selection step; objects that are deemed point sources in the $V$ and
$R$ images of NGC~4157 are plotted with filled circles; objects deemed
extended are plotted with open squares.  The number of sources
remaining after this step was 271 in NGC~2683, 179 in NGC~4157, 275 in
NGC~3556, 262 in NGC~3044, and 245 in NGC~7331.

\subsection{Photometry}

Before doing photometry of the objects in the source lists, we
computed individual aperture corrections for each image of each galaxy
by measuring the light from 10$-$20 bright stars
within a series of apertures from 1 to 6 times the average FWHM of the
image.  The aperture corrections are listed in Table~\ref{table:aper
corr} and represent the mean difference between the total magnitude of
the bright stars and the magnitude within the aperture with radius one
FWHM.  Photometry with an aperture of radius equal to the average FWHM
of the images was then performed for each of the sources that remained
after the extended source cut.  Calibrated $BVR$ magnitudes were
derived for each source by taking the instrumental magnitude and
applying the appropriate aperture correction and photometric
calibration coefficients.  Corrections for Galactic extinction,
derived from the reddening maps of \citet{schlegel98}, were also
applied to produce final $BVR$ magnitudes.  Galactic extinction
corrections are listed in Table~\ref{table:ext corr}.

\subsection{Color Selection}
\label{section:color selection}

The last step in selecting GC candidates is to choose from the list of
point sources the objects with $V$ magnitudes and $BVR$ colors
consistent with their being GCs at the distance of the host galaxy.
This was executed following the same basic steps for all the
galaxies. First, objects with $M_V$ $<$ $-$11 (assuming the distance
moduli given in Table~\ref{table:galaxy properties}) were removed from
the lists.  Then, if an object had a $B-V$ color and error that put it
in the range 0.56 $<$ $B-V$ $<$ 0.99, it was selected.  (This $B-V$
range corresponds to [Fe/H] of $-$2.5 to 0.0 for Galactic GCs; Harris
1996.)  Finally, if the objects had $V-R$ colors and errors that put
them within a specified distance from the relation between $B-V$ and
$V-R$ for Milky Way GCs, they were selected as GC candidates.

In practice, small refinements to this basic set of selection criteria
were applied to produce the final list of GC candidates.  When the
colors of the objects that pass the extended source cut are plotted in
the $BVR$ color-color plane, a marked overdensity of sources in the
region of the plane occupied by Galactic GCs is usually obvious.  We
therefore adjusted the selection criteria slightly to ensure that all
the objects within these ``overdensities'' were selected as GC
candidates.  Furthermore, because there are relatively few GC
candidates around the spiral galaxies (typically $<$100), we examined
each of the objects with magnitudes and colors anywhere close to those
of Galactic GCs, to confirm that we were not missing real GCs or
(conversely) including likely contaminants in the GC candidate
samples.

For NGC~2683, NGC~3044, and NGC~4157, we accepted all objects that
were within 3-$\sigma$ above or below the $V-R$ vs. $B-V$ line for
Milky Way GCs (where $\sigma$ is the scatter in the $V-R$ vs. $B-V$
relation). In addition, in NGC~4157, we accepted two sources that were
physically located very near the disk of the galaxy and had colors
that put them just outside the $BVR$ selection box, in the direction
of the reddening vector.  For NGC~7331, we used a 2-$\sigma$ criterion
for the $BVR$ color selection in order to exclude several likely
contaminants, and we accepted three objects near the galaxy disk with
$BVR$ colors indicating they were probably reddened GCs.  For
NGC~3556, the overdensity of point sources in the globular cluster
region of the $BVR$ color-color plane was weighted toward the blue
side of the $V-R$ vs. $B-V$ relation for Milky Way GCs.  Therefore for
this galaxy, we selected objects that were 1.5-$\sigma$ above (redder
than) the $V-R$ vs. $B-V$ relation, and 3-$\sigma$ below (bluer than)
the relation.  After examining the objects selected in the $BVR$
color-color plane, we decided to also apply a $V$ magnitude cut for
two of the galaxies, in order to eliminate what appeared to be a
significant number of faint background objects that had not been
removed in the extended source cut.  We excluded objects with $V$
$>$23.0 in NGC~2683, and $V$ $>$ 23.5 in NGC~3556 which is,
respectively, approximately 0.9 mag and 0.4 mag past the peak of the
GC luminosity function in those galaxies.  After all of these
magnitude and color selection criteria were applied, the final numbers
of GC candidates found in NGC~2683, NGC~3556, NGC~4157, and NGC~7331
was 41, 50, 37, and 37, respectively.

NGC~3044 was a special case.  When we plotted the 262 point sources
detected around this galaxy in the $BVR$ color-color plane, it was
immediately apparent that there was no overdensity of objects in the
part of the plane occupied by GCs.  We nevertheless applied a typical
set of color selection criteria --- i.e., $B-V$ in the usual range and
$V-R$ within 3$\sigma$ above or below the expected value for Galactic
GCs --- and created a list of 35 possible GC candidates, with
$V$~$=$~20.9$-$24.4.  These objects were spread uniformly over the
field of view of the WIYN images, rather than being strongly clustered
around the galaxy, as is typical for the other spiral galaxy GC
systems we have surveyed.  Only 2 of the 35 objects with GC-like
magnitudes and colors were located within a projected radius of 2.2
arc~minutes (15 kpc, assuming the 23~Mpc distance) from the galaxy
center.  (In the Milky Way, more than 80\% of the catalogued GCs have
projected radial distances of 15~kpc or less; Harris 1996.)  This
suggests that the GC system of NGC~3044 was not clearly detected with
our WIYN observations.  Assuming that this galaxy's distance modulus
is 31.83 and that its GCLF peaks at $M_V$ $=$ $-$7.33 like the Milky
Way GCLF (Ashman \& Zepf 1998), the GCs in the luminous half of the
GCLF should have $V$ magnitudes in the range 20.8$-$24.5.  The survey
images are typically 50\% complete at $B$, $V$, and $R$ magnitudes of
24$-$25 (see Section~\ref{section:completeness}), so many such
luminous GCs should be detectable in the images.  It may be that this
galaxy has a $S_N$ significantly lower than those of the Milky Way and
the other spiral galaxies in our survey (and thus has very few
luminous GCs), that the galaxy actually lies somewhat further away
than 23~Mpc, and/or that the magnitude limits of the images and our
selection techniques prevent us from detecting the GCs that are there.
In any case, because no convincing GC candidates were detected in
NGC~3044, no further analysis steps were executed for this galaxy.

\begin{figure}
\plotone{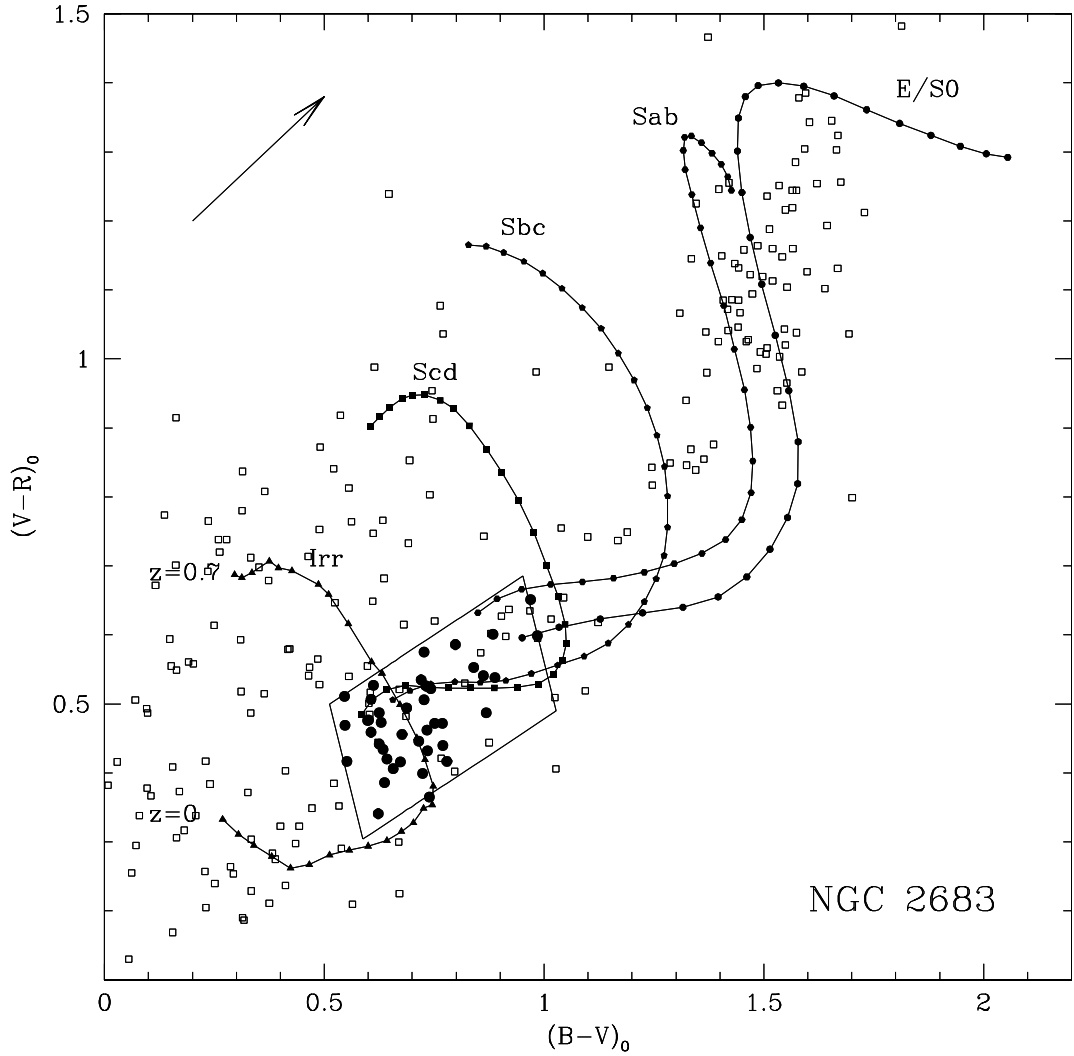}
\caption{\normalsize Color selection of GC candidates in NGC~2683.  Open squares
  are the 271 point sources detected in all three filters; filled
  circles are the final set of 41 GC candidates.  For reference, the
  locations in the $BVR$ color-color plane of galaxies of various
  types are shown as tracks the galaxies would follow with increasing
  redshift.  RZ01 details how the tracks were produced.  A reddening
  vector of length $A_V = 1$ mag appears in the upper left-hand
  corner.}
\label{fig:bvr n2683}
\end{figure}

\begin{figure}
\plotone{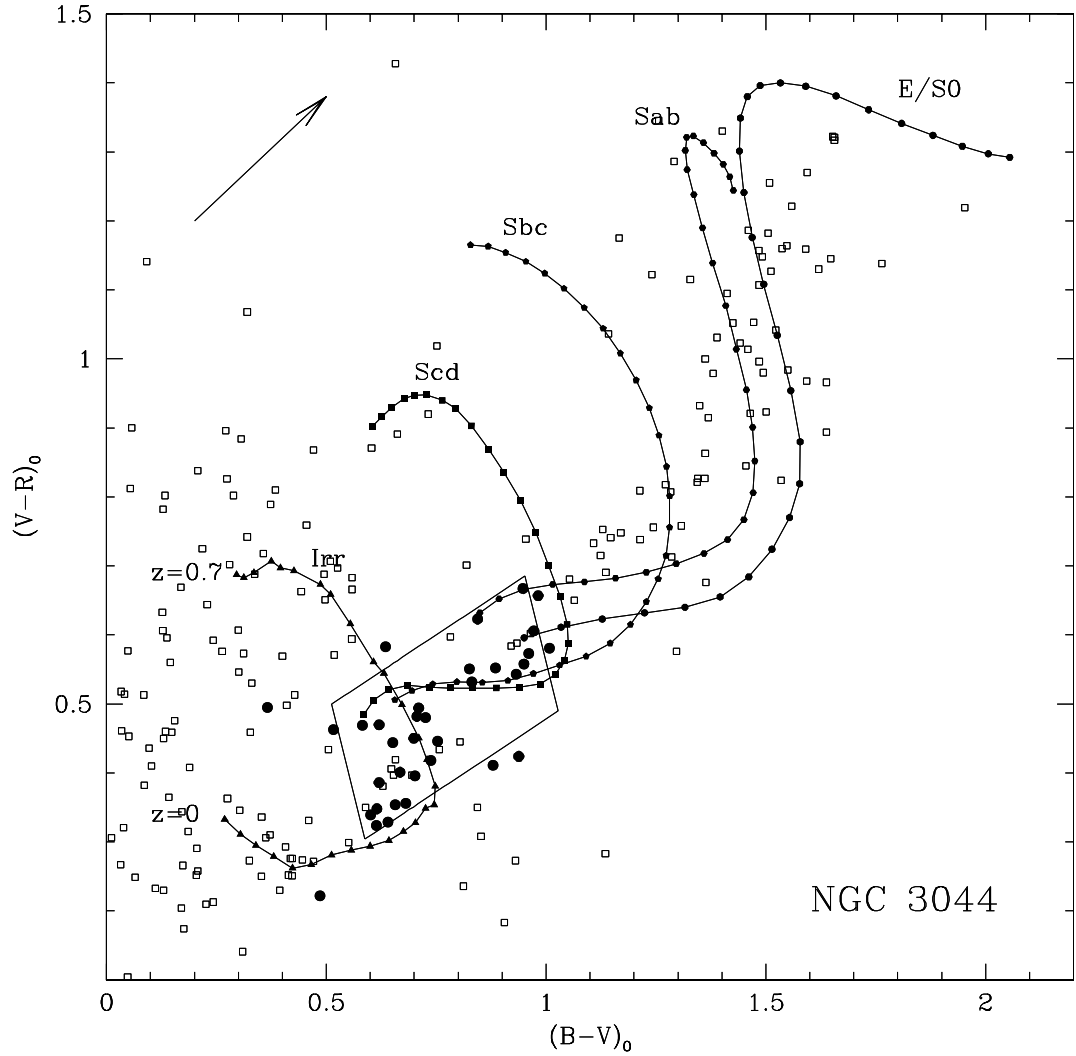}
\caption{\normalsize Color selection of GC candidates in NGC~3044.  Open squares
  are the 262 point sources detected in all three filters; filled
  circles are 35 sources with $V$ magnitudes and $BVR$ colors like
  globular clusters.  Note that only these 35 objects are spread
  uniformly over the entire WIYN image, and in
  Section~\ref{section:color selection} we argue that we have not
  convincingly detected this galaxy's GC system.  }
\label{fig:bvr n3044}
\end{figure}

\begin{figure}
\plotone{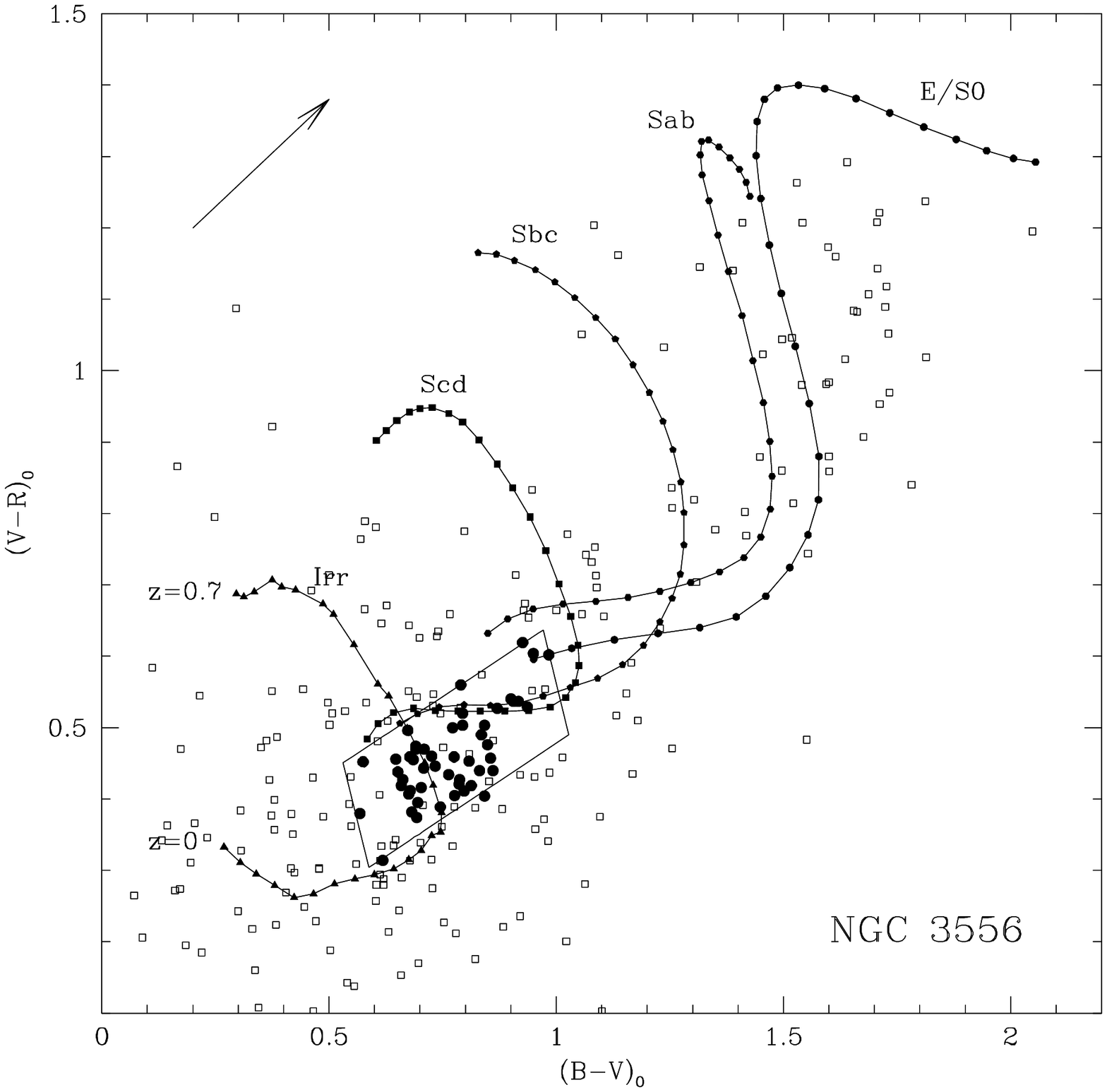}
\caption{\normalsize Color selection of GC candidates in NGC~3556.  Open squares
  are the 275 point sources detected in all three filters; filled
  circles are the final set of 50 GC candidates.}
\label{fig:bvr n3556}
\end{figure}

\begin{figure}
\plotone{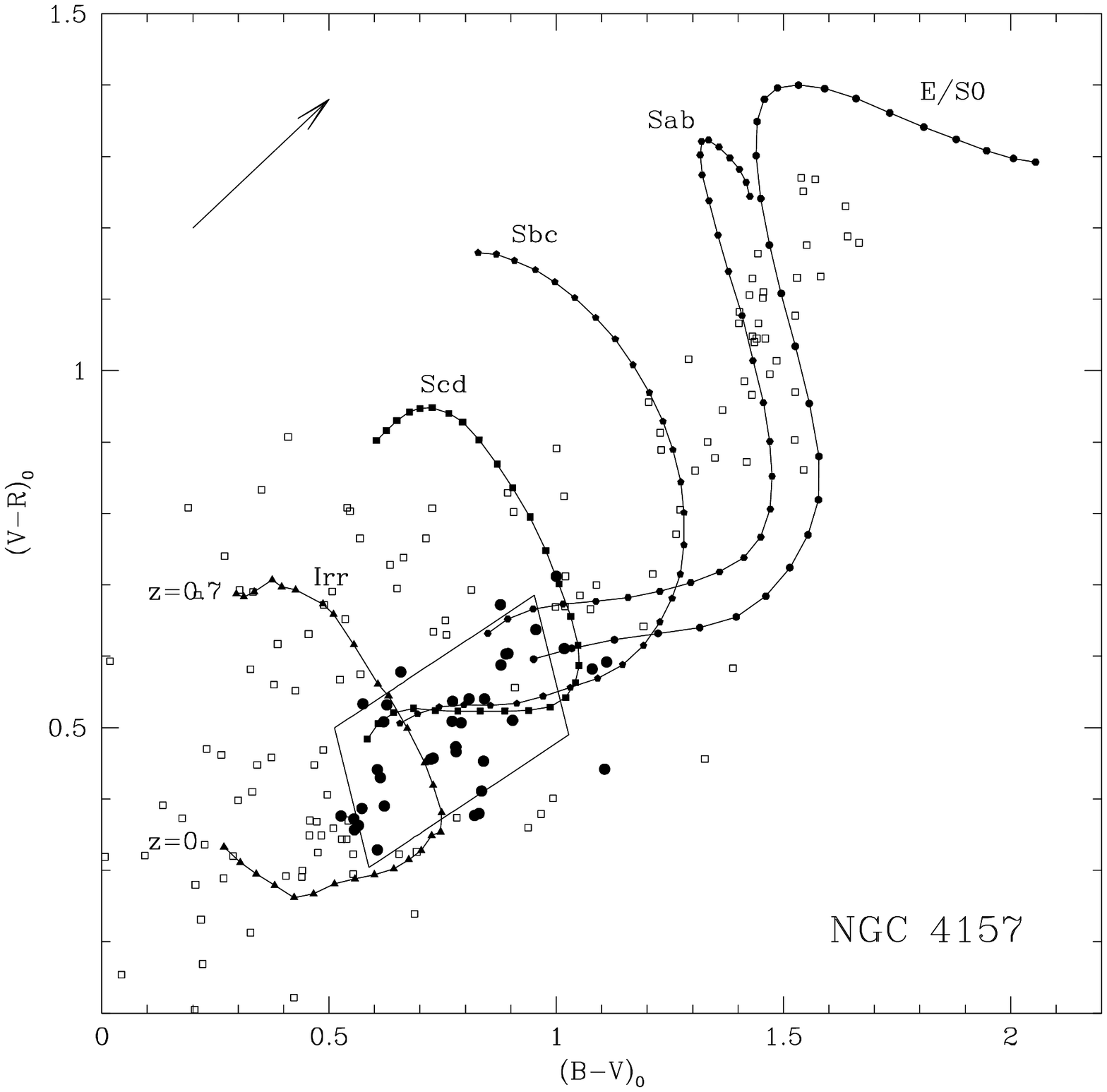}
\caption{\normalsize Color selection of GC candidates in NGC~4157.
  Open squares are the 171 point sources detected in all three
  filters; filled circles are the final set of 37 GC candidates.}
\label{fig:bvr n4157}
\end{figure}

\begin{figure}
\plotone{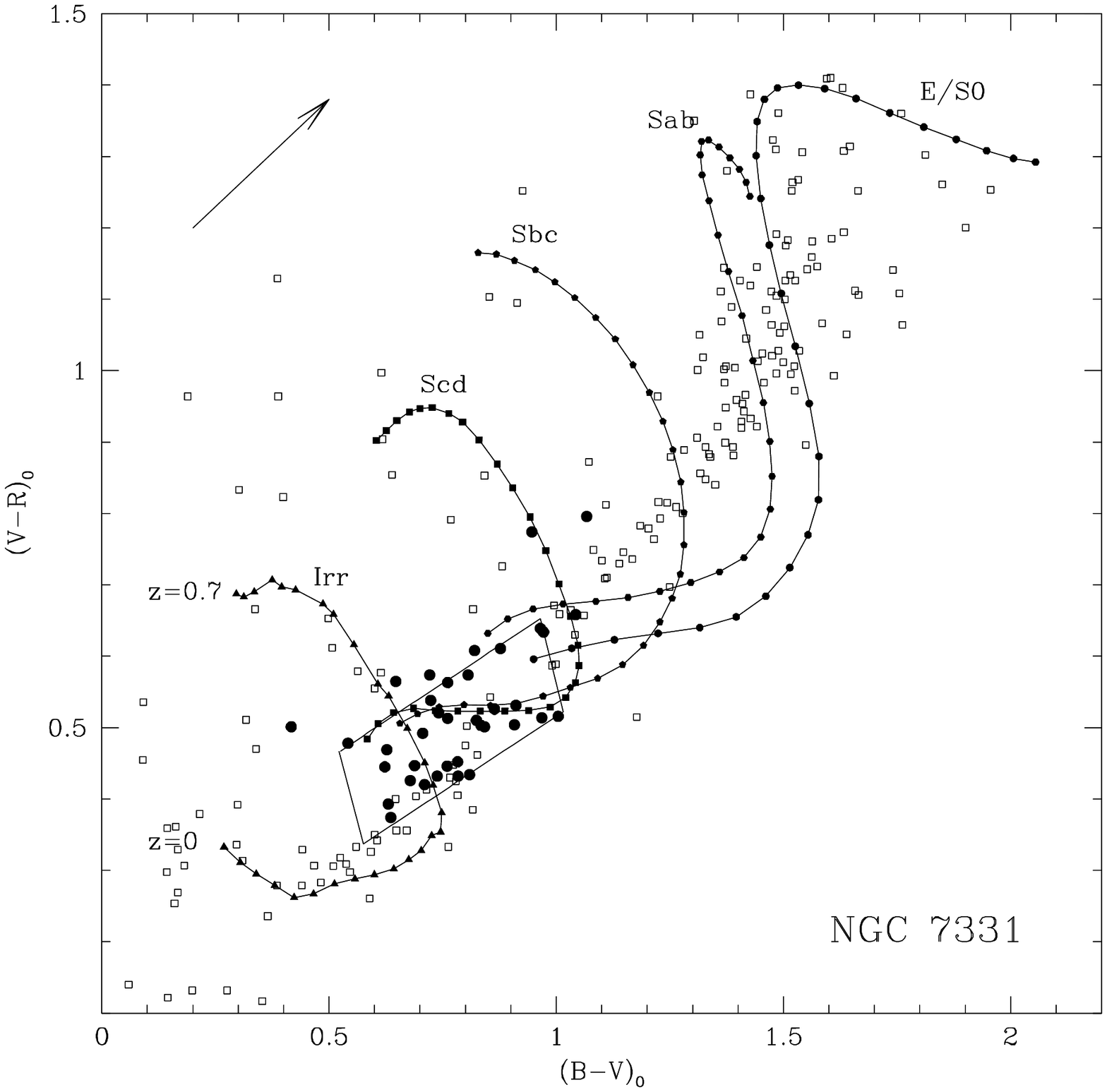}
\caption{\normalsize Color selection of GC candidates in NGC~7331.
  Open squares are the 245 point sources detected in all three
  filters; filled circles are the final set of 37 GC candidates.}
\label{fig:bvr n7331}
\end{figure}

Figures~\ref{fig:bvr n2683} through \ref{fig:bvr n7331} illustrate the
results of the color selection; objects appearing as point sources in
the WIYN images are shown as open squares, and filled circles are the
final selected GC candidates.  We include the $BVR$ color-color plots
for NGC~3044 for completeness.  Note that because of the $V$ magnitude
criteria applied, some objects within the color selection boxes are
not selected as GC candidates. For illustrative purposes, the figures
show the expected locations of galaxies of different morphological
types, at different redshifts.  The ``galaxy tracks'' were produced by
taking template galaxy spectra for early- to late-type galaxies,
shifting the spectra to simulate moving the galaxies to redshifts
between 0 and 0.7, and then calculating their $BVR$ colors (see RZ01
for details). The galaxy tracks simply show that late-type, low- to
moderate-redshift galaxies have $BVR$ colors similar to GCs, so not
every GC candidate in the samples at this stage is a real GC; some may
be background galaxies.  (Section~\ref{section:contamination} details
our efforts to quantify the amount of contamination in the GC
candidate lists.)  

\begin{figure}
\plotone{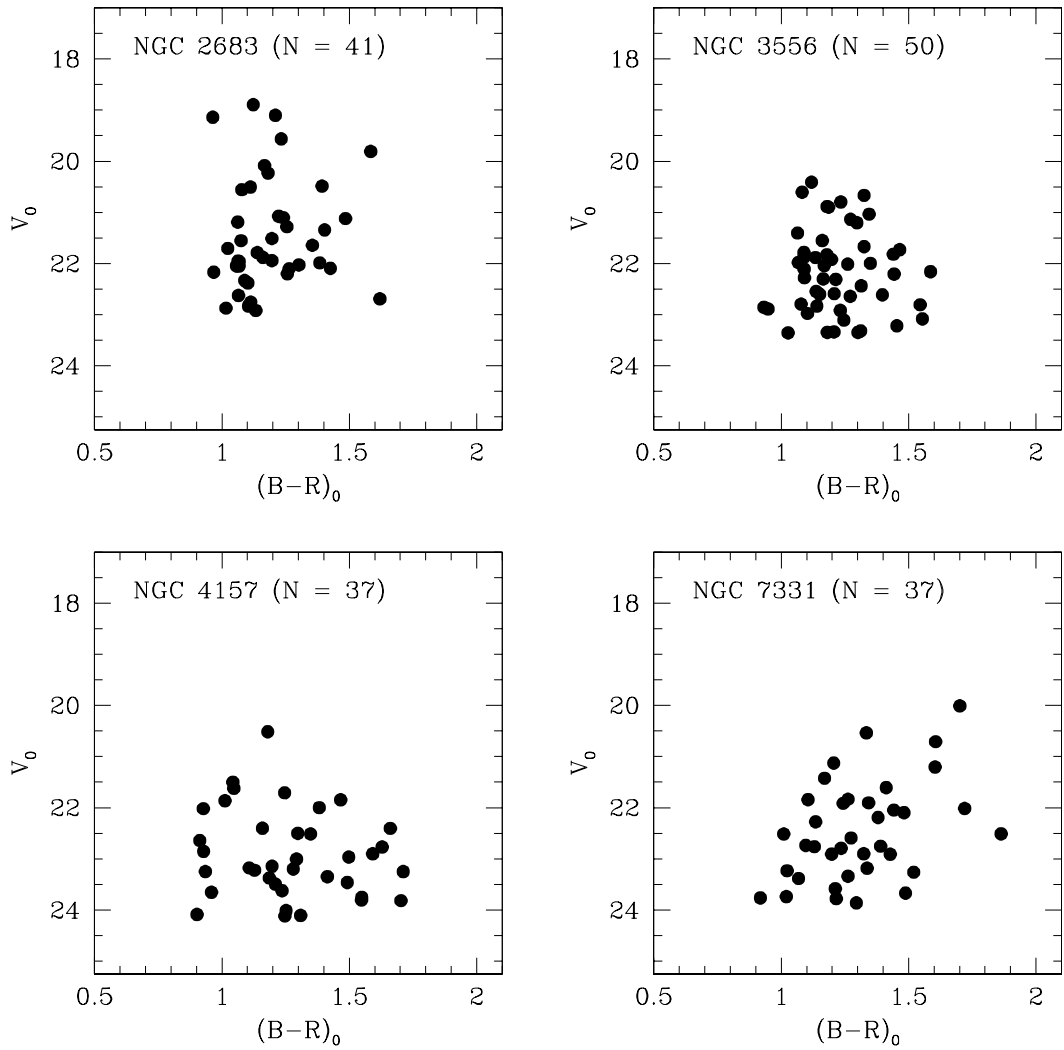}
\caption{\normalsize $V$ versus $B-R$ color-magnitude diagrams for the GC
  candidates found in each of the four galaxies in which the GC system
  was detected.  The magnitudes and colors have been corrected for
  Galactic extinction in the direction of the target galaxies, but not
  for absorption internal to the galaxies.}
\label{fig:four cmds}
\end{figure}

Figure~\ref{fig:four cmds} shows color-magnitude diagrams for the GC
candidates in the four galaxies in which the GC system was detected.
The $V$ magnitudes of the GC candidates are plotted versus their $B-R$
colors. The final numbers of GC candidates found in the galaxies are
marked on the plots. 

\subsection{Completeness Testing}
\label{section:completeness}

A series of completeness tests was done to determine the point-source
detection limits of the WIYN images of each galaxy.  We began by
adding artificial point sources with magnitudes within 0.2~mag of a
particular mean value to each of the $B$, $V$, and $R$ images.  The
number of artificial sources depended on the size of the images: 200
sources were added to the Minimosaic images and 50 were added to the
S2KB images (which cover one-fourth the area of the Minimosaic
frames).  Next, the same detection steps performed on the original
images were performed on the images containing the artificial stars,
and the fraction of artificial stars recovered in the detection
process was recorded. The process was repeated 25$-$30 times for each
image, incrementing the mean magnitude of the artificial stars by
0.2-mag each time, so that the completeness was calculated over a
range of 5$-$6 magnitudes for each filter, for each galaxy.
Table~\ref{table:completeness} lists the 50\% completeness limits for
the three filters for each galaxy.
 
\subsection{Quantifying and Correcting for Contamination}
\label{section:contamination}

Some fraction of the objects chosen as GC candidates are actually
contaminating objects --- that is, foreground stars or background
galaxies that have $BVR$ magnitudes and colors like globular clusters.
We used a combination of techniques to estimate the amount of
contamination from non-GCs that existed in our samples, so that we
could correct for this contamination in subsequent analysis steps.

\subsubsection{Contamination Estimate Based on the Asymptotic Behavior of the Radial
  Profile} 
\label{section:asymptotic contam}

Our first step for this set of galaxies was to use the observed radial
distribution of GC candidates to help estimate the contamination
level.  First, an initial radial profile of the GC system of each
galaxy was constructed by assigning the GC candidates to a series of
annuli, each 1$\arcm$ in width.  (More details about the construction
of the radial distributions are given in
Section~\ref{section:profiles}.) The effective area of each annulus
(the region where GCs could be detected) was determined and used to
calculate a surface density of GC candidates for the annulus.  The
resultant surface density profile for each galaxy's GC system followed
the general shape expected for GC systems, but rather than going to
zero in the outer regions, the profiles decreased until reaching a
constant positive value in the last few annuli.  We assumed this
constant surface density was due to contaminating objects (stars and
galaxies).  We calculated the weighted mean surface density of objects
in these outer annuli and took this as an estimate of the
contamination level of the GC candidate lists.

For NGC~2683, the initial radial profile created at this step
flattened to a constant value in the outer six annuli (of a total of
nine annuli).  The mean surface density of objects in these outer six
annuli is 0.226$\pm$0.026 arcmin$^{-2}$.  For NGC~3556, a constant
surface density of GC candidates was present in the outer four of
eight annuli.  Here the level was 0.093$\pm$0.003~arcmin$^{-2}$.  For
NGC~4157, the situation was more complicated.  The initial radial
distribution of GC candidates showed typical behavior in the inner few
annuli (beginning at some maximum surface density value and then
monotonically decreasing with increasing radius) but then showed a
``bump'' of increased surface density in two adjacent bins in the
outer profile, at $\sim$30~kpc from the galaxy center.  This feature
in the radial profile is caused by a group of GC candidates in the
halo of the galaxy and is discussed in detail in
Section~\ref{section:profiles}.  The objects may be {\it bona fide}
GCs in the galaxy's outer halo, or a distant, background cluster of
galaxies masquerading as GCs, or just a chance superposition of
several otherwise-unrelated GC candidates.  In any case, to estimate
the surface density of contaminants, we removed the seven
closely-grouped GC candidates responsible for the inflated surface
density in those two outer bins, and calculated the mean surface
density of GC candidates in the outer five (of nine total) annuli.
The estimated surface density of contaminants from this analysis is
0.181$\pm$0.059~arcmin$^{-2}$.  Finally, for NGC~7331, the images
covered less area around the galaxy, so the radial coverage of the GC
system was reduced.  The initial radial profile of GC candidates
flattened in the outer two (of five) annuli and the average surface
density in these annuli is 0.690$\pm$0.220~arcmin$^{-2}$.

\subsubsection{Estimating Stellar Contamination from a Galactic Star
  Counts Model}
\label{section:star contam}

We used the Galactic structure model code of \citet{mendez96} and
\citet{mendez00} to yield an independent estimate of the level of
contamination from Galactic stars in the GC candidate lists.  Given
specific values for variables such as the Galactocentric distance of
the Sun and the proportions of stars in the Galaxy halo, disk, and
thick disk, the code outputs the predicted number of Galactic stars
with $V$ magnitudes and $B-V$ colors in a user-specified range, in a
given area of the sky.  We ran the model for each of our galaxy
targets, adjusting the range of $V$ magnitudes of the Galactic stars
to match the $V$ range of the GC candidates.  The model-predicted
surface density of stars with $V$ magnitudes and $B-V$ colors like the
GC candidates in the directions NGC~2683, NGC~3556, NGC~4157, and
NGC~7331 were 0.067~arcmin$^{-2}$, 0.040~arcmin$^{-2}$,
0.051~arcmin$^{-2}$, and 0.155~arcmin$^{-2}$, respectively.  (The
stellar surface density in the direction of NGC~7331 is large compared
to the other galaxies because this galaxy is at $-$20 degrees Galactic
latitude, significantly closer to the Galactic plane than the other
galaxies, which have Galactic latitudes between $+$40 and $+$65
degrees.)  The surface density values were relatively insensitive to
the choice of model parameters for the Galaxy.  The surface density
values range from $\sim$20$-$40\% of the total surface density of
contaminants estimated from the asymptotic behavior of the radial
profile.

\subsubsection{Estimating the Contamination from Background Galaxies
  with HST Data}
\label{section:hst contam}

Archival {\it HST} imaging data\footnote{Based on observations made
with the NASA/ESA {\it Hubble Space Telescope}, obtained from the data
archive at the Space Telescope Science Institute.  STScI is operated
by AURA, under NASA contract NAS 5-26555.} were available for some of
the target spiral galaxies.  Because $HST$ can resolve many extended
objects that appear as point sources in ground-based imaging,
determining whether the WIYN GC candidates are actually galaxies gives
us another estimate of the contamination in the WIYN data.

We downloaded all of the available archival $HST$ images taken in
broadband filters of the targeted galaxies; all of these images were
taken with the Wide Field and Planetary Camera~2 (WFPC2).  We found
images that covered portions of the WIYN pointings of NGC~2683,
NGC~3556, and NGC~7331, but no data for NGC~4157.  The data sets we
analyzed are summarized in Table~\ref{table:hst data}.  The table
lists: proposal ID; target name (either the name of the galaxy or
``Any'', which indicates that the images were taken by WFPC2 while
another $HST$ instrument was being used for the primary science);
total exposure time; distance of the observation from the galaxy
center; and filter.  ``On-the-fly'' calibration was applied to the
images before they were retrieved from the archive.  The STSDAS task
GCOMBINE was used to combine individual exposures of the same
pointing. The WIYN GC candidates were then located in the WFPC2
images.  We used the method of \citet{kundu99}, who measured the flux
from GC candidates in apertures of radius 0.5 pixels and 3 pixels and
then calculated the ratio of counts in the large and small apertures.
Objects that are extended (and therefore galaxies) have count ratios
much larger than those of point sources, since relatively more of
their light is contained within the 3-pixel aperture.  We confirmed
the results from the count-ratio method with visual inspection.

For NGC~2683, we analyzed a pointing 0.7~arcmin from the galaxy
center, taken in the 606W filter, that turned out to contain none of
the WIYN GC candidates.  We also analyzed a WFPC2 pointing in the 814W
filter, centered 1.9~arcmin from the center of the galaxy.  Thirteen
of the WIYN GC candidates appear in the WFPC2 field and one of these
is an extended object rather than a real GC.  The area covered by the
WFPC2 image is 5.527 arcmin$^{2}$, so one estimate of the number
density of background galaxies (from these admittedly small-number
statistics) is 0.181~arcmin$^{-2}$.

For NGC~3556, there was a WFPC2 pointing in the 606W filter, located
0.4~arcmin from the galaxy center.  Five of the WIYN GC candidates
appear in the combined WFPC2 image; none is extended.

NGC~7331 had many WFPC2 images available in the HST archive, largely
because this galaxy was a target for the HST Cepheid Key Project
\citep{hughes98}.  We made combined images from multiple observations
of two different pointings located 3.5$\arcmin$ and 5$\arcmin$ from
the galaxy (see Table~\ref{table:hst data}).  Only one of these
pointings contained WIYN GC candidates, however; the other pointing
happened to coincide with a small area of the WIYN frames that
contained no GC candidates.  Two of the WIYN GC candidates are located
within the WFPC2 pointing at 3.5~arcmin from the galaxy center;
neither object is extended.

\subsubsection{Final Contamination Correction}
\label{section:final contam}

For the spiral galaxies here, 
we will adopt the contamination levels
estimated from the asymptotic behavior of the radial profile, and
take the Galactic star counts models and HST data analysis as checks
on these estimates.  NGC~2683 is the one galaxy for which we have
independent estimates of the contamination from both stars (from the
star counts model) and galaxies (from $HST$ data).  Adding these two
numbers together (0.181~arcmin$^{-2}$ $+$ 0.067~arcmin$^{-2}$) gives
the same number within the errors as the total contamination estimate
from the radial profile (0.226$\pm$0.026~arcmin$^{-2}$).  Also, as
noted in Section~\ref{section:star contam}, the stellar contamination
from the Galactic star counts model was always lower than the total
contamination estimated from the radial profile, which makes sense if
one assumes that galaxies also contribute to the contamination level
in the GC candidate samples.

We took the number density of contaminating objects for each galaxy
given in Section~\ref{section:asymptotic contam} and used it to
calculate the expected fraction of contaminants at each annulus in the
radial profile, for use in subsequent steps.  First we multiplied the
number density of contaminants by the effective area of the annulus.
Dividing this number by the total number of GC candidates in the
annulus then yielded the fraction of contaminating objects for that
annulus.

\subsection{Determining the GCLF Coverage}

The observed GC luminosity function (GCLF) was constructed for each of
the four galaxies by assigning the $V$ magnitudes of the GC candidates
to bins of width 0.3 mag.  The radially-dependent correction described
in Section~\ref{section:final contam} was used to correct the LF data
for contamination.  For example, if a GC candidate was located
2~arcmin from the galaxy center and the contamination fraction was
expected to be 20\% at that radius, then 0.8 was added to the total
number of objects in the appropriate $V$ magnitude bin.  The LF was
also corrected for completeness, by computing the total completeness
of each $V$ bin (calculated by convolving the completenesses in all
three filters, as detailed in RZ01) and dividing the number of GCs in
that bin by the completeness fraction.

We assumed the intrinsic GCLF of the spiral galaxies was a Gaussian
function with a peak absolute magnitude like that of the Milky Way
GCLF, $M_V$ $=$ $-$7.3 \citep{az98}.  If one applies the distance
moduli in Table~\ref{table:galaxy properties}, this $M_V$ translates
to peak apparent magnitudes of $V$ $=$ 22.1, 23.1, 23.5, and 23.3, for
NGC~2683, NGC~3556, NGC~4157, and NGC~7331, respectively.  We fitted
Gaussian functions with the appropriate peak apparent magnitude and
dispersions of 1.2, 1.3, and 1.4 mag to the corrected LF data.  Bins
with less than 45\% completeness were excluded from the fitting
process.  In a few instances we also excluded one or more bins with
very low numbers (e.g., bins containing less than one GC candidate)
that were causing the normalization to be too low to fit well to the
surrounding bins.  The fraction of the theoretical GCLF covered by the
observed LF was calculated for each galaxy.  The mean fractional
coverage (averaged for the three different dispersions) and error for
NGC~2683, NGC~3556, NGC~4157, and NGC~7331 were 0.64$\pm$0.02,
0.53$\pm$0.02, 0.513$\pm$0.001, and 0.51$\pm$0.02, respectively.

Finally, we experimented with changing the bin size of the LF data and
quantified how this affected the final value of the fractional GCLF
coverage.  We found that changing the bin size produced a change in
the mean fractional coverage of 4$-$7\%.  This uncertainty is included
in the final errors (i.e., the error on specific frequency) on
quantities discussed in Section~\ref{section:total numbers}.

\section{Results}
\label{section:results}

\subsection{Radial Distributions of the GC Systems}
\label{section:profiles}

We constructed radial profiles of the galaxies' GC systems by binning
the GC candidates into a series of 1$\arcm$-wide annuli according to
their projected radial distances from the galaxy centers.  The inner
parts of the spiral galaxy disks had been masked out (see
Section~\ref{section:source detection}) because GCs could not be
reliably detected there; thus the positions of the radial bins were
adjusted so that the inner radius of the first annulus started just
outside this masked central region.  An effective area --- the area in
which GCs could be detected, excluding the masked portions of the
galaxy, masked regions around saturated stars, and parts of the
annulus that extended off the image --- was computed for each annulus.
We corrected the number of GCs in each annulus for contamination (by
applying the radially dependent contamination correction described in
Section~\ref{section:final contam}) and for GCLF coverage.  The final
radial distribution of the GC system was then produced by dividing the
corrected number of GCs in each annulus by the effective area of the
annulus.  The errors on the GC surface density values for each annulus
include uncertainties on the number of GCs and contaminating objects.
Tables~\ref{table:profile n2683} through \ref{table:profile n7331}
give the final radial distributions of the GC systems for the four
galaxies; the radial profiles are plotted in Figures~\ref{fig:profile
n2683} through \ref{fig:profile n7331}.  The projected radii shown in
the tables and figures are the mean projected radii of the unmasked
pixels in each annulus. Note that because a contamination correction
has been applied to the surface density of GCs in each radial bin,
some of the outer bins have negative surface densities.

\begin{figure}
\plotone{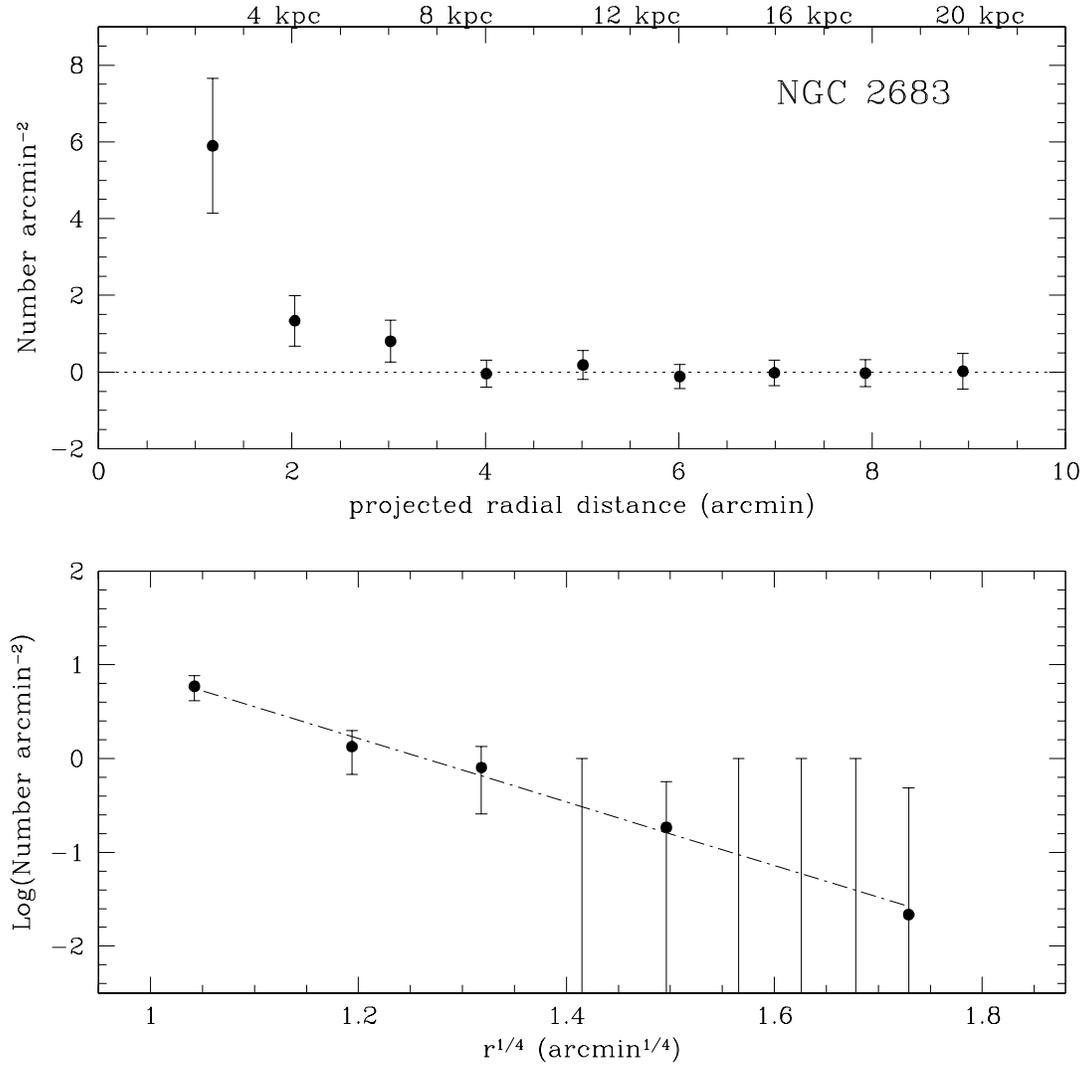}
\caption{\normalsize Radial distribution of GCs in NGC~2683, plotted as surface
  density vs.\ projected radial distance (top) and as the log of the
  surface density vs.\ $r^{1/4}$ (bottom).  The horizontal line in the
  top panel indicates a surface density of zero.  The dashed line in
  the bottom panel is the best-fit de~Vaucouleurs law.  The data have
  been corrected for contamination, areal coverage, and magnitude
  incompleteness, as described in Section~\ref{section:profiles}.}
\label{fig:profile n2683}
\end{figure}

\begin{figure}
\plotone{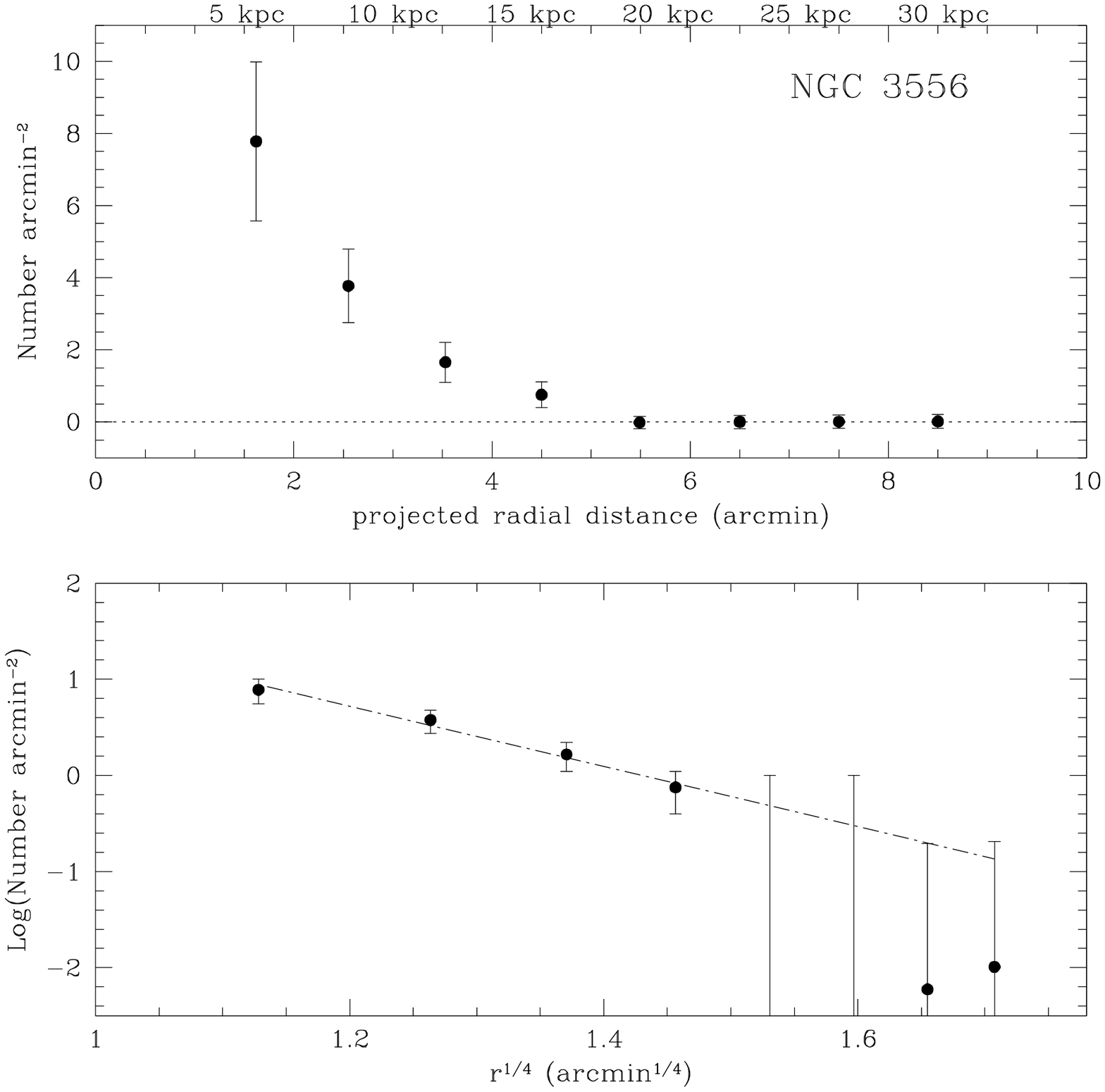}
\caption{\normalsize Radial distribution of GCs in NGC~3556, plotted the same way
  as in Fig. \ref{fig:profile n2683}.}
\label{fig:profile n3556}
\end{figure}

\begin{figure}
\plotone{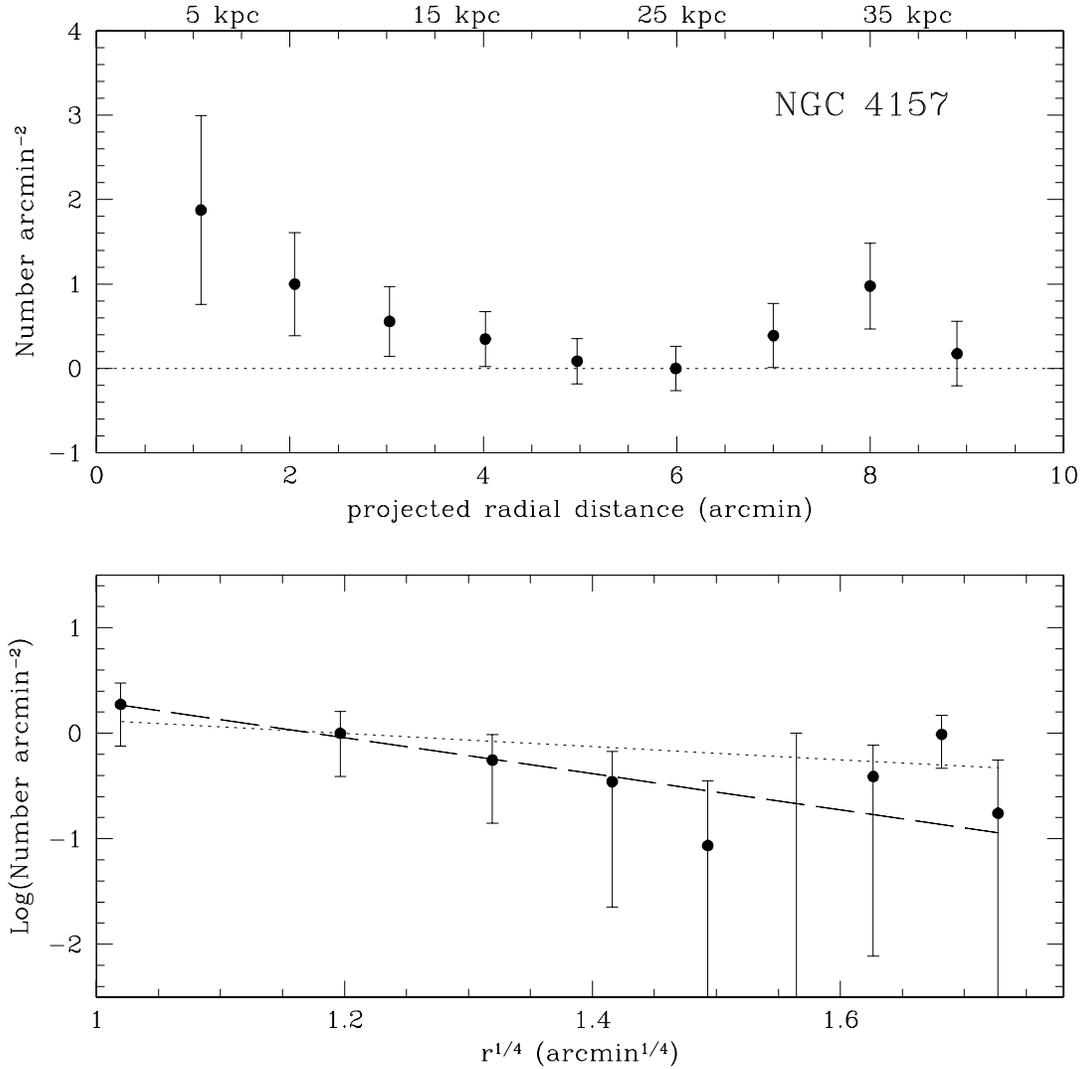}
\caption{\normalsize Radial distribution of GCs in NGC~4157, showing the surface
  density vs. projected radius in the top panel and the log of the
  surface density vs. $r^{1/4}$ in the bottom panel. There is a spike
  in the surface density in the 7$-$8$\arcm$ bins, due to the presence
  of seven loosely-grouped objects that may or may not be real GCs in
  NGC~4157's halo.  The dotted line in the bottom panel is the
  best-fit deVaucouleurs law with those seven objects included in the
  data; the dashed line is the deVaucouleurs law fit with the seven
  objects removed.}
\label{fig:profile n4157}
\end{figure}

\begin{figure}
\plotone{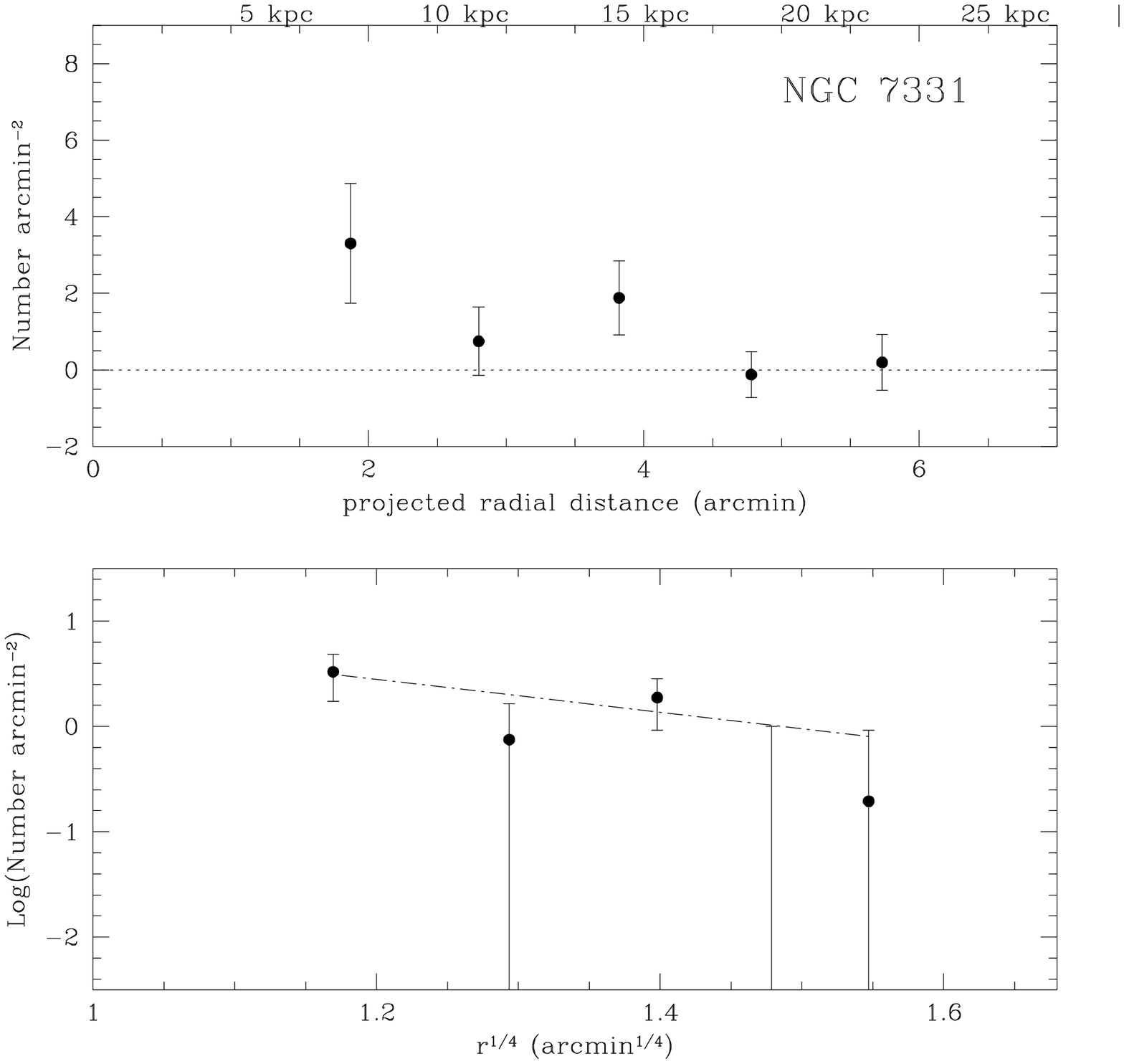}
\caption{\normalsize Radial distribution of GCs in NGC~7331, plotted the same way
  as in Fig. \ref{fig:profile n2683}.}
\label{fig:profile n7331}
\end{figure}

We fitted power laws of the form log~$\sigma_{\rm GC}$ $=$ $a0$ $+$
$a1$~log~$r$ and deVaucouleurs laws of the form log~$\sigma_{\rm GC}$
$=$ $a0$ $+$ $a1$~$r^{1/4}$ to the radial distributions.  In all cases
the $\chi^2$ values were nearly the same for both the power law and
deVaucouleurs law fits, so we list the coefficients for both functions
in Table~\ref{table:profile coefficients}.  The top panels of
Figures~\ref{fig:profile n2683} through \ref{fig:profile n7331} show
the surface density of GCs versus projected radius and the bottom
panels show the log of the surface density versus $r^{1/4}$, with the
best-fit deVaucouleurs law plotted as a dashed line.

The GC system radial profiles of each of the galaxies have slightly
different appearances, but in all cases the GC surface density
decreases to zero within the errors before the last data point.  This
suggests that we have observed the full radial extent of the galaxies'
GC systems, which is crucial for a reliable determination of the total
number of GCs in the system.  For NGC~2683, the surface density is
consistent with zero by a radius of 4$\arcm$, or $\sim$9~kpc.  For
NGC~3556, which is a much more luminous spiral galaxy, the GC surface
density goes to zero at 5.5$\arcm$, or $\sim$20~kpc.

NGC~7331 has the lowest inclination of any spiral galaxy in our
survey.  (In addition to the current set of galaxies, the survey also
includes NGC~7814, published in RZ03, and NGC~891 and NGC~4013, which
are not yet published.)  Because the galaxy has $i$ $\sim$75 degrees
(rather than the $i$ $\sim$ 80$-$90 degrees of the other target
galaxies), we had to mask out a relatively large portion of its inner
radial region, because we could not reliably detect GCs against the
background of the galaxy's dusty spiral disk.  Consequently the
innermost point in the galaxy's radial profile (Fig.~\ref{fig:profile
n7331}) is centered at $>$7 kpc from the galaxy center.  In general,
the spiral galaxy GC systems we have studied are fairly concentrated
toward the galaxy center (as is true for the Milky Way).  This seems
to be the case for NGC~7331 as well: outside of the masked galaxy disk
region, we barely detect NGC~7331's GC system before the data points
in the radial profile flatten to a constant surface density (which we
take to be the contamination level in the data; see
Section~\ref{section:contamination}).  In the final version of the
radial profile shown in Figure~\ref{fig:profile n7331}, the surface
densities in the first three radial bins are positive (although the
errors on the surface density at 3$\arcm$ make its lower limit barely
consistent with zero).  Then the GC surface density goes to zero
within the errors in the fourth and fifth radial bins. The fourth
radial bin is centered at 4.8$\arcm$ (18~kpc), so we take this as the
approximate radial extent of this galaxy's GC system.

NGC~4157 was another special case.  The radial profile shows the
expected behavior in the inner regions of the GC system: the surface
density of GCs decreases monotonically with increasing radius until it
is consistent with zero in the radial bins centered at 5$\arcm$ and
6$\arcm$.  However the surface density then increases to positive
values in the 7$\arcm$ and 8$\arcm$ radial bins, before returning to
zero in the last radial bin at 9$\arcm$.  This ``bump'' in the outer
profile is caused by the presence of seven GC candidates with
projected distances of $\sim$30~kpc from the galaxy center.  The seven
candidates appear in two groups: one close group with three objects at
$r$ $=$ 31.0$-$31.7~kpc, and a group of four objects that are somewhat
more spread out, with four objects at $r$ $=$ 27$-$32~kpc. These
grouped objects may be real GCs in NGC~4157's halo, or a distant
background group or cluster of galaxies.  Alternatively, the objects
may just appear near each other by chance.  

We thought it possible that some or all of the seven GC candidates in
question might be associated with a dwarf galaxy that is being
accreted into NGC~4157's halo.  For this reason we obtained deep,
broadband images of NGC~4157 with the WIYN Minimosaic in March 2005.
The combined images have a total integration time of 7.5 hours and
reach a surface brightness level of $V$ $>$27, but we found no
evidence for a faint dwarf galaxy anywhere in the vicinity of those
seven GC candidates.

We fitted deVaucouleurs and power laws to both the original version of
the radial profile of NGC~4157's GC system and to a profile with the
seven GC candidates at $\sim$30~kpc removed.  Both fits are shown in
Figure~\ref{fig:profile n4157} and listed in Table~\ref{table:profile
n4157}.  With the seven ``extra'' GC candidates removed, the GC
surface density is consistent with zero within the errors by 5$\arcm$,
or 20~kpc.

\subsection{Radial Extent of the GC Systems of the Survey Galaxies}
\label{section:extents}

\begin{figure}
\plotone{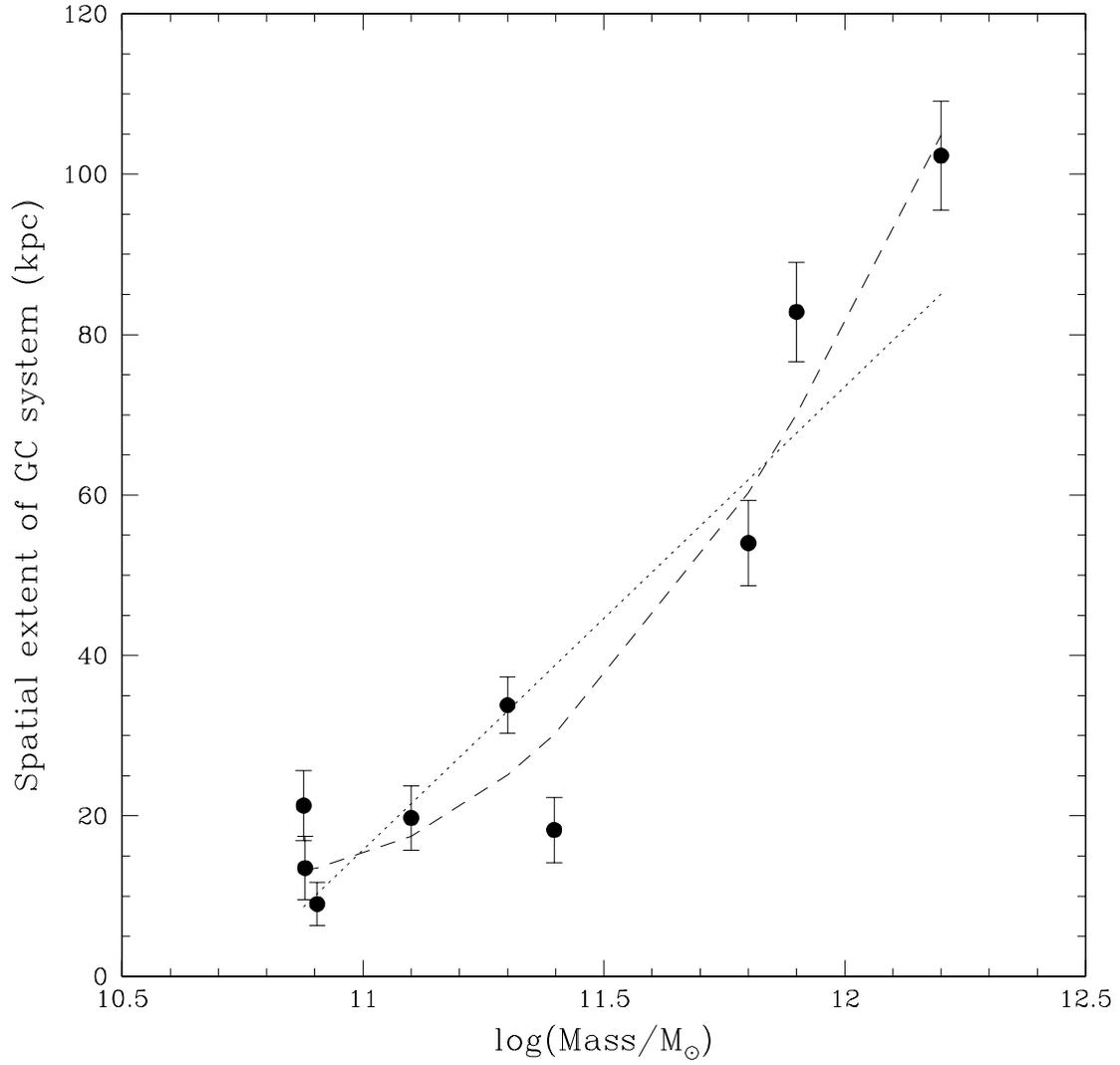}
\caption{\normalsize Radial extent of the globular cluster system in kiloparsecs
  versus the log of the galaxy mass in solar masses for nine
  elliptical, S0, and spiral galaxies included in our wide-field GC
  system survey to date.  The best-fit line and second-order
  polynomial are shown as dotted and dashed lines, respectively.}
\label{fig:extent}
\end{figure}

With (including these four spiral galaxies) a total of nine galaxies
from the wide-field survey now analyzed, we can look for trends in the
GC system properties of the overall sample.  One quantity that we
derive from the radial profiles of the GC systems is an estimate of
the radial extent of the systems.  For the survey, we take the radial
extent to be the point at which the surface density of GCs in the
final radial profile becomes consistent with zero (within the errors
on the surface density) and stays at zero out to the radial limit of
the data.  Figure~\ref{fig:extent} shows the radial extent in
kiloparsecs of the GC systems of the nine survey galaxies analyzed to
date, plotted against the log of the host galaxy stellar mass.  To
compute the galaxy masses, we combined $M^T_V$ for each galaxy with
the mass-to-light ratios given in \citet{za93}: $M/L_V$ $=$ 10 for
elliptical galaxies (NGC~3379, NGC~4406, and NGC~4472) $M/L_V$ $=$ 7.6
for S0 galaxies (NGC~4594), $M/L_V$ $=$ 6.1 for Sab$-$Sb galaxies
(NGC~2683, NGC~4157, NGC~7331, and NGC~7814) and 5.0 for Sbc$-$Sc
galaxies (NGC~3556).  The errors on the radial extent values in
Figure~\ref{fig:extent} were calculated by taking into account the
errors on the distance modulus assumed for each galaxy, along with the
errors on the determination of the radial extent itself (which we took
to be equal to the width of one radial bin of the spatial profile).
Note also that for the spiral galaxy NGC~4157, we derived the radial
extent from the version of the radial profile with the seven ``extra''
GC candidates discussed in Section~\ref{fig:extent} removed.  The
galaxy stellar masses and estimated radial extents are listed in
Table~\ref{table:extents}.

Figure~\ref{fig:extent} shows that, as might be expected, more massive
galaxies generally have more extended GC systems, although with a fair
amount of scatter in the relation. We fitted a line and a second-order
polynomial to the data in the figure; the best-fit line and curve are,
respectively: 

\begin{equation}
y = ((57.7\pm3.7)~x) - (619\pm41)
\end{equation}

and 

\begin{equation}
y = ((45.7\pm9.5)~x^2) - ((985\pm217)~x) + (5320\pm1240)
\end{equation}

\noindent where $x$ is the radial extent in kpc and $y$ is
$log(Mass/M_{\odot})$.  One useful application of the data and
best-fit relations is determining how much radial coverage is needed
in order to observe all or most of the GC system of a particular
galaxy.  For example, the field of view of the $HST$ Advanced Camera
for Surveys (ACS) is 3.37$\arcm$ on a side, which means that the
maximum radial range observable with this instrument is 2.38$\arcm$ if
the galaxy is positioned at the center of the detector.  At the
approximate distance of the Virgo Cluster ($\sim$17~Mpc), this
translates to a maximum projected radial distance of $\sim$12~kpc.
Therefore if one intends to observe the full radial extent of the GC
system of a galaxy placed at the center of the HST ACS field,
Figure~\ref{fig:extent} indicates that one is limited to galaxies with
stellar mass $log(M/M_{\odot})$ $<$ 11 (which is less massive than the
Milky Way Galaxy). One ACS field would include $\sim$50\% of the
radial extent of a GC system of a Virgo Cluster galaxy with
$log(M/M_{\odot})$ in the range 11.1$-$11.2.  As explained in
Section~\ref{section:introduction} of this paper and in other papers
from the survey (e.g., RZ01, RZ03, RZ04), deriving accurate global
values for the properties of a GC system requires that one observe
most or all of the radial extent of the system.

\subsection{Total Number and Specific Frequency of GCs}
\label{section:total numbers}
 
The number of GCs in each galaxy can be derived by integrating the
deVaucouleurs profiles fitted to the radial distributions (see
Section~\ref{section:profiles}) out to some outer radial limit.  Since
the radial distributions are corrected for magnitude incompleteness,
missing spatial coverage, and contamination from non-GCs, the result
is a final estimate of the total number of GCs in the system
($N_{GC}$). We chose the outer radius of integration to be the point
in the radial distribution at which the surface density of GCs equals
zero within the errors, and then remains consistent with zero for the
remainder of the data points.  Note that for NGC~4157, we integrated
the deVaucouleurs profile that was fitted to the radial profile with
the seven ``extra'' GC candidates located in the 7$-$8$\arcm$ bins
removed.  The outer radius of integration in that case was 5$\arcm$,
because with those seven objects removed, the GC surface density is
consistent with zero beginning with the 5$\arcm$ bin out to the last
bin at 9$\arcm$.

Given that we could not observe some portion of the inner galaxy ---
close to the spiral disk --- for all four target galaxies, we also had
to make assumptions about the number of GCs and/or the shape of the GC
radial profile within that region.  The projected outer radial
boundary of the unobserved region ranged from 0.5$-$1.3$\arcm$, which
translates to 1$-$5~kpc given the distances to the target galaxies.
We assumed four different possibilities for the behavior of the GC
systems in these inner regions: (1) that the same proportion of GCs
was located within the region as in the Milky Way GC system; (2) that
the proportion of GCs missing was like the GC system of NGC~7814
(which we did observe to small radius, with HST WFPC2; see RZ03); (3)
that the best-fit deVaucouleurs law profile continued all the way to
$r$~$=$~0; and (4) that the inner part of the GC radial distribution
was flat (i.e., the GC surface density in the unobserved region
equalled the value in the first radial bin of the observed profile).
Adding the number of GCs in the observed part of the system
(calculated from integrating the deVaucouleurs profile over the radial
range of the data) to the number of GCs in the inner region (given
these various assumptions) yielded a range of values for the total
number of GCs in each galaxy's system.  We took the mean of this range
of values as the final estimate of $N_{GC}$.

The luminosity- and mass-normalized numbers of GCs in a galaxy are
useful quantities to calculate and compare among galaxies.  The
specific frequency, $S_N$ was defined by \citet{hvdb81} as

\begin{equation}
{S_N \equiv {N_{GC}}10^{+0.4({M_V}+15)}}
\end{equation}

The $M_V$ values assumed for the galaxies are those given in
Table~\ref{table:galaxy properties}.  An alternative quantity, $T$,
was introduced by \citet{za93} and is sometimes preferred to $S_N$
because it normalizes the number of GCs by the stellar mass of the
galaxy ($M_G$) rather than $V$-band magnitude:

\begin{equation}
T \equiv \frac{N_{GC}}{M_G/10^9\ {\rm M_{\sun}}}
\end{equation}

\noindent To calculate $M_G$, we again combined $M^T_V$ for each
galaxy with mass-to-light ratios from \citet{za93}.  The total
numbers, $S_N$ and $T$ values for each galaxy's GC system are given in
Table~\ref{table:total numbers}.

To calculate the errors on $N_{GC}$ and the specific frequencies $S_N$
and $T$, we took into account the following sources of uncertainty:
(1) the variation in the total number of GCs, depending on what was
assumed for the spatial distribution of the unobserved inner portion
of the GC system; (2) the variation in the calculated coverage of the
GCLF, depending on the assumed intrinsic GCLF function and how the
luminosity function data were binned; and (3) Poisson errors on the
number of GCs and the number of contaminating objects.  For NGC~4157,
we also estimated the uncertainty in $N_{GC}$ due to the group of
seven GC candidates in the galaxy's halo, and whether these were real
GCs (and should therefore be included in the total number) or
contaminants.  Errors on the specific frequencies $S_N$ and $T$ also
include uncertainties in the total galaxy magnitudes: we assumed that
the internal extinction correction (and thus the galaxy magnitude)
could be uncertain by as much as 0.3 mag, which corresponds to 3$-$4
times the error on $V^0_T$ given in RC3 \citep{devauc91}. Individual
errors from the above sources were added in quadrature to calculate
the final errors on $N_{GC}$, $S_N$ and $T$; the errors are given in
Table~\ref{table:total numbers}.

The GC system of one of our targets, NGC~2683, was studied previously
by \citet{harris85}.  They used photographic data from the
Canada-France-Hawaii telescope to identify $\sim$100 GC candidates and
estimated that the total number of GCs in NGC~2683 is 321$\pm$108.
Combining this number with our assumed $M_V^T$ value of $-$20.5 yields
a specific frequency $S_N$ of 2.0$\pm$0.7.  This is 2.5 times larger
than our measured $S_N$ value (0.8$\pm$0.4). It is not unusual for our
survey data to yield smaller $N_{GC}$ and $S_N$ values than previous
studies; of six galaxies (including NGC~2683) with $S_N$ values
already in the literature, four have previously-published $S_N$ values
significantly larger than those we derive (RZ01, RZ03, RZ04).  Our
smaller $S_N$ values are probably due to a combination of factors,
e.g., lower contamination levels (due to source selection in multiple
filters and higher resolution data) and more accurate radial
distributions yielding better-determined total numbers of GCs.

\begin{figure}
\plotone{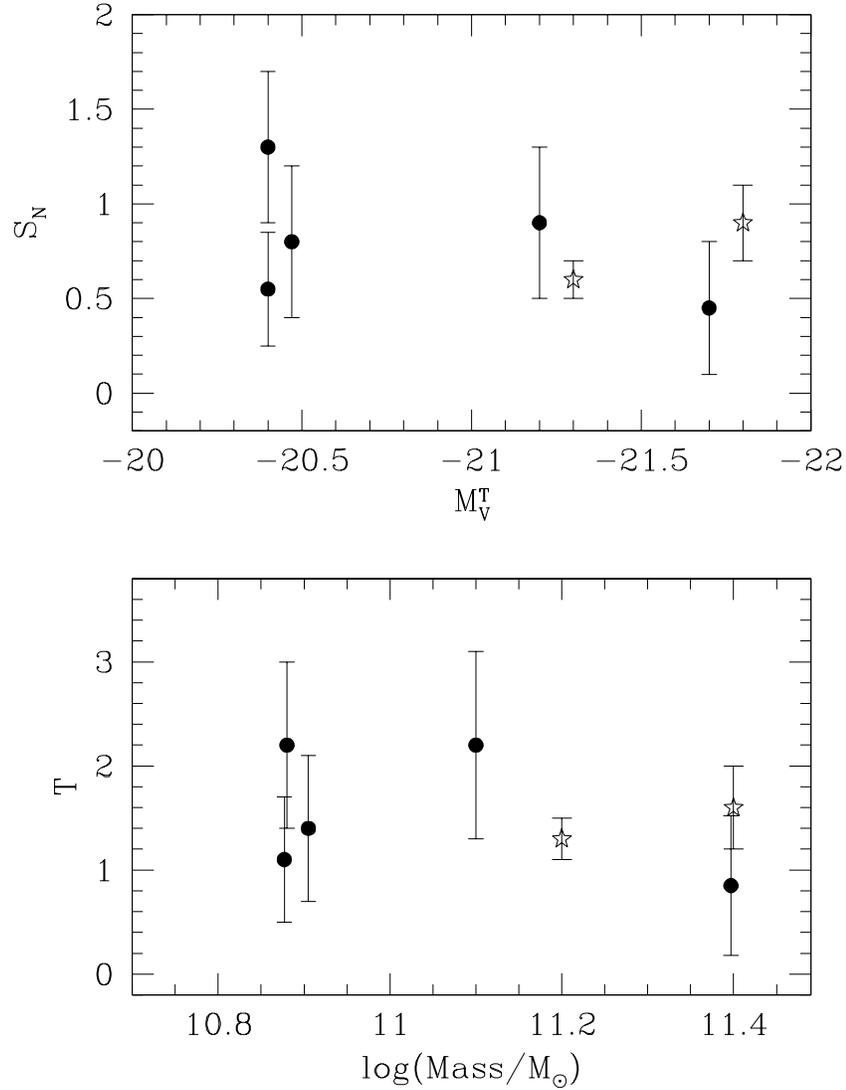}
\caption{\normalsize Luminosity-normalized specific frequency ($S_N$) and
  mass-normalized specific frequency ($T$) for the five spiral
  galaxies (four presented here and one presented in RZ03) analyzed
  for our wide-field GC system survey, plotted with filled circles.
  Specific frequencies of the GC systems of the Milky Way (smallest
  error bars) and M31 are plotted with open stars. The specific
  frequencies for the targeted spiral galaxies fall within a modest
  range and are consistent with the values for the Milky Way and M31.}
\label{fig:spec freq}
\end{figure}

One of the objectives of our wide-field CCD survey was to determine
whether the two spiral galaxies with the most thoroughly studied GC
systems, the Milky Way and M31, are typical of their galaxy class in
terms of the properties of their GC systems.  Figure~\ref{fig:spec
freq} addresses this question by comparing the luminosity- and
mass-normalized specific frequencies of the Milky Way and M31 with
those of the spiral galaxies from our survey.  The specific
frequencies of the five galaxies we have analyzed are plotted with
filled circles.  Besides the four spiral galaxies presented in this
paper, this includes the Sab galaxy NGC~7814 (RZ03). The five spiral
galaxies we have analyzed have morphological types of Sab (N=1
galaxy), Sb (N $=$3), and Sc (N$=$1) and stellar masses in the range
$log(M/M_{\odot})$ $=$ 10.9$-$11.4.
Open stars in Figure~\ref{fig:spec freq} indicate $S_N$ and $T$ for
the Milky Way (smaller error bars) and M31.  The Milky Way has
$N_{GC}$ $\sim$ 180, $S_N$ = 0.6$\pm$0.1, and $T$ $=$ 1.3$\pm$0.2
\citep{az98}.  M31 has $\sim$450~GCs, $S_N$ $=$ 0.9$\pm$0.2 and $T$
$=$ 1.6$\pm$0.4 (Ashman \& Zepf 1998; Barmby et al.\ 2000).  The seven
spiral galaxies in the figure show a fairly small range of specific
frequency values, with modest scatter (note that $S_N$ can range from
less than zero to $>$10 for giant galaxies; Ashman \& Zepf 1998).  The
weighted mean $S_N$ and $T$ values for the GC systems of the five
spiral galaxies in the survey are 0.8$\pm$0.2 and 1.4$\pm$0.3,
respectively.  These values fall between, and are consistent with, the
$S_N$ and $T$ values for the GC systems of the Milky Way and M31,
which suggests that the spiral galaxies we are most familiar with are
indeed representative of the GC systems of other spiral galaxies of
similar mass, at least in terms of the relative number of GCs.  The
mean $N_{GC}$ of the five spiral galaxies we have surveyed to date is
170$\pm$40.

\citet{goud03} analyzed the GC systems of six nearly edge-on spiral
galaxies with HST WFPC2 optical imaging data.  For five of the
galaxies, they had a single WFPC2 observation positioned near the
galaxy center; for the sixth galaxy, they had two WFPC2 fields on each
side of the disk. Given the distances to their target galaxies, the
WFPC2 pointings provided radial coverage of the GC systems out to
(typically) $\sim$5$-$15~kpc.  To calculate total numbers and specific
frequencies of GCs in the target galaxies, they follow a technique
from \citet{kissler99} and compare the numbers of GCs detected in the
WFPC2 data with the numbers that would be detected at the same spatial
location in the Milky Way GC system if it were observed under the same
conditions (i.e., at the same distance and projected onto the sky in
the same manner).  The mean $S_N$ value for the five Sab$-$Sc spiral
galaxies studied by Goudfrooij et al.\ is 0.96$\pm$0.26 and the mean
$T$ value is 2.0$\pm$0.5.

\citet{chandar04} used HST WFPC2 imaging to study the GC systems of
five low-inclination spiral galaxies.  They had very limited spatial
coverage of the galaxies' GC systems: usually the data consisted of
$\sim$1$-$4 WFPC2 pointings located within the inner 5$\arcm$
($<$14~kpc) of each galaxy's disk.  To correct for their missing
spatial coverage, \citet{chandar04} used a technique similar to the
one used by \citet{kissler99} and \citet{goud03}: they compared the
locations of GCs in their observed fields to the analogous locations
and fields in the Milky Way GC system, if it were observed face-on.
They calculated a scale factor (equal to the ratio of the number of
GCs in the analogous Milky Way region to the number of GCs detected in
their observed fields) and applied it to the total number of GCs in
the target galaxy.  As Goudfrooij et al.\ and Chandar et al.\ both
note, the implicit assumption in this method is that GC systems of
other spirals have the same spatial distributions as that of the Milky
Way.  For the five spiral galaxies in the Chandar et al.\ study, the
average $S_N$ is 0.5$\pm$0.1 and average $T$ is 1.3$\pm$0.2.

The mean $S_N$ and $T$ values found by Goudfrooij et al.\ and Chandar
et al.\ are consistent, within the errors, with the average $S_N$ and
$T$ values we derive from observing the majority of the radial extent
of the GC systems.  This is perhaps not entirely unexpected, because
our mean $S_N$ and $T$ values are in line with the corresponding
values for the Milky Way, and the {\it HST} studies necessarily had to
assume similarity with the Milky Way GC system in order to calculate
their total numbers and specific frequencies.

\subsection{Number of Blue (Metal-Poor) GCs Normalized by Galaxy Mass}

Another specific objective of the overall wide-field GC system survey
is to test a prediction of AZ92, who suggested that elliptical
galaxies and their GC populations can form from the merger of two or
more spiral galaxies.  In the AZ92 model, the GCs associated with the
progenitor spiral galaxies form a metal-poor GC population in the
resultant elliptical, and a second, comparatively metal-rich
population of GCs is formed during the merger itself.  For simple
stellar populations older than $\sim$1$-$2~Gyr, broadband colors
primarily trace metallicity, with bluer colors corresponding to lower
metallicities and red to higher metallicities (see, e.g., Ashman \&
Zepf 1998).  AZ92 therefore predicted that giant elliptical galaxies
should show at least two peaks in their broadband color distributions,
due to the presence of the metal-poor (blue) GCs associated with the
original spiral galaxies and metal-rich (red) GCs formed in star
formation triggered by the merger. Bimodal GC color distributions have
subsequently been observed in many elliptical galaxies (e.g., Zepf \&
Ashman 1993, Kundu \& Whitmore 2001; Peng et al.\ 2006).  (A detailed
discussion of the exact relationship between color and metallicity for
old stellar populations in different broadband colors --- and whether
the bimodal color distributions of elliptical galaxy GC systems really
are due to the presence of distinct GC subpopulations --- is beyond
the scope of this paper.  Thorough discussion of these issues is given
in, e.g., \citet{zepf07}, \citet{strader07}, and \citet{kz07}.)  A
consequence of the AZ92 scenario is that the mass-normalized specific
frequencies of blue, metal-poor GCs in spiral and elliptical galaxies
should be about the same.  Our survey data allow us to calculate this
quantity, $T_{\rm blue}$, and compare it for galaxies of different
morphological types.  We define $T_{\rm blue}$ as

\begin{equation}
T_{\rm blue} \equiv \frac{N_{GC}(\rm blue)}{M_G/10^9\ {\rm M_{\sun}}}
\end{equation}

\noindent where $N_{GC}(\rm blue)$ is the number of blue GCs and $M_G$
is the stellar mass of the host galaxy, calculated by combining
$M^T_V$ with $M/L_V$, as described in Section~\ref{section:extents}.

Estimating the proportion of blue GCs in the early-type galaxies from
the survey is fairly straightforward because of the large numbers
(hundreds to thousands) of GC candidates detected in these galaxies.
We make this estimate by first constructing a sample of GC candidates
that is at least 90\% complete in all three of our imaging filters
($B$, $V$, and $R$), creating a GC color distribution from the
complete sample, and then running a mixture-modeling code to fit
Gaussian functions and estimate the proportion of GCs in the blue and
red peaks.  The code we use, called KMM \citep{abz94}, requires at
least 50 objects (given the typical color separation between
metal-poor and metal-rich GC systems) to produce reliable results.

Calculating $T_{\rm blue}$ for the spiral galaxies is more uncertain
because of poor statistics: we typically detect tens of GC candidates
in the galaxies, and end up with very few objects in the sample of
candidates that is complete in $B$, $V$, and $R$.  For NGC~2683,
NGC~3556, NGC~4157, and NGC~7331, there were 38, 31, 7, and 26 objects
in the complete sample used to construct the GC color
distribution. Because we are interested in testing whether the blue GC
populations in elliptical galaxies could have originated in spiral
galaxies, we define as ``blue'' those GCs with $B-R$ $<$ 1.23, the
typical location of the separation between the blue and red GC
populations in elliptical galaxies (RZ01, RZ04).  The percentage of GC
candidates with $B-R$ $<$ 1.23 in the complete samples constructed for
NGC~2683, NGC~3556, NGC~4157, and NGC~7331 were 63\%, 55\%, 57\%, and
31\%.  We took these percentages to be lower limits on the percentage
of blue GCs for the overall system, since presumably some GCs we
detect might be reddened due to internal extinction. Since it seems
unlikely that {\it all} of the GC candidates in the complete sample
are reddened and belong in the blue category, we assumed that, at
most, 70\% of the candidates might be blue.  We based this number on
the GC systems of the Milky Way and M31, which both show two peaks in
their color and metallicity distributions.  Roughly 70\% of the Milky
Way GCs lie in the metal-poor peak \citep{harris96}.  For M31, the
proportion is similar.  \citet{barmby00} estimate from both
photometric and spectroscopic metallicities that $\sim$66\% of M31's
GCs lie in the metal-poor peak.  \citet{perrett02} publish
spectroscopic estimates of [Fe/H] for $>$200 M31 GCs and estimate that
77\% are metal-poor.

We converted these lower- and upper-limit blue percentages into
$T_{\rm blue}$ values (with an associated error) for each galaxy by
multiplying them by the galaxy's total $T$ value and error.  We then
averaged the lower- and upper-limit $T_{\rm blue}$ values, and their
errors, and took that as the final estimate of $T_{\rm blue}$ for each
galaxy.  These values are given in Table~\ref{table:total numbers}.
We note that preliminary $T_{\rm blue}$ values for NGC~2683, NGC~3556,
and NGC~4157 were included in \citet{rzs05}; the values given in the
current paper are the same except that the final calculated errors are
slightly larger for NGC~2683 and NGC~4157.

The $T_{\rm blue}$ values for the nine galaxies analyzed so far as
part of this wide-field GC system imaging survey are shown in
Figure~\ref{fig:tblue}.
This is an updated version of a figure initially shown in
\citet{rzs05}.  The figure now includes seven early-type galaxies and
seven spiral galaxies: four early-type galaxies and five spiral
galaxies from our survey (RZ01, RZ03, RZ04, and the current paper);
the Milky Way (Zinn 1985; Ashman \& Zepf 1998); M31 (Ashman \& Zepf
1998; Barmby et al.\ 2000; Perrett et al.\ 2002); and three elliptical
galaxies from the literature (NGC~1052 from Forbes et al.\ 2001;
NGC~4374 from Gomez \& Richtler 2004; NGC~5128 from Harris, Harris, \&
Geisler 2004) that meet our criteria for inclusion in the figure.
(Namely, that at least 50\% of the radial extent of the GC system is
observed, and enough information that the total number of GCs and blue
fraction could be estimated; see \citet{rzs05} for details.)  In the
figure, circles denote cluster elliptical galaxies, squares mark field
E/S0 galaxies, and triangles are spiral galaxies in the field.  Filled
symbols are used for our survey data, the Milky Way, and M31; data
from other studies are shown with open symbols.  (The curves in the
figure are discussed below.)

One immediately sees from Figure~\ref{fig:tblue} that the $T_{\rm
  blue}$ values for the spiral galaxies we surveyed are relatively
  uncertain compared to the early-type galaxies, as a direct result of
  the poor number statistics described earlier.  Even with these
  uncertainties, it is apparent that the typical $T_{\rm blue}$ value
  for spiral galaxies is smaller than that of the more massive cluster
  elliptical galaxies.  The weighted mean $T_{\rm blue}$ for the
  cluster ellipticals is 2.3$\pm$0.2, compared to 0.9$\pm$0.1 for the
  spiral galaxies.  This suggests that some other mechanism ---
  besides the straightforward merging of spiral galaxies envisioned by
  AZ92 --- is needed to create massive cluster elliptical galaxies and
  their GC populations.  On the other hand, the $T_{\rm blue}$ values
  of three of the four field E/S0 galaxies in the figure are
  comparable to those of the spiral galaxies, which suggests that
  merging spiral galaxies and their GC systems together is sufficient
  to account for the metal-poor GC populations of some field E/S0
  galaxies.  Similar conclusions were made in RZ03, RZ04, and
  \citet{rzs05} and still hold here, with (now) finalized estimates of
  $T_{\rm blue}$ for seven spiral galaxies.  (We should note here that
  previous authors have compared {\it total} GC specific frequency
  values for elliptical and spiral galaxies, and reached basically the
  same result.  For example, \citet{harris81} compares $S_N$ for
  elliptical galaxies in the Virgo cluster and the field to $S_N$ for
  the Milky Way, M31, and two other spiral galaxies.  \citet{harris81}
  concludes that the merger of disk galaxies with $S_N$ $<$3 would
  produce a new elliptical with $S_N$ in the range $\sim$1 to 3, which
  is in line with $S_N$ for some field ellipticals, but much lower
  than $S_N$ for Virgo cluster ellipticals.)

Also apparent in Fig.~\ref{fig:tblue} is a general trend of increasing
$T_{\rm blue}$ value with increasing host galaxy stellar mass: more
massive galaxies tend to have proportionally more metal-poor GCs.  We
first discussed this trend in detail in \citet{rzs05} and noted there
that it is consistent with a biased, hierarchical galaxy formation
scenario such as that suggested by \citet{santos03}. In this picture,
the first generation of GCs form at high redshift during the initial
stages of galaxy assembly.  The GCs are metal-poor because they form
from relatively unenriched gas.  This first epoch of GC and baryonic
structure formation is then temporarily suppressed at $z$
$\sim$10$-$15; \citet{santos03} suggests that the suppression is
triggered by the reionization of the Universe. Massive galaxies in
high-density environments are associated with higher peaks in the
matter density distribution, and therefore began their collapse and
assembly process first.  The result of such ``biasing'' is that
massive galaxies had assembled a larger fraction of their eventual
total mass by the suppression redshift.  A larger fraction of their
baryonic mass could therefore participate in the formation of the
first generation of GCs and as a result, more massive galaxies end up
with relatively larger $T_{\rm blue}$ values compared to less massive
galaxies.  \citet{santos03} assumes that the break from baryonic
structure formation is fairly short-lived ($<$1~Gyr).  During this
period, stellar evolution continues to enrich the intergalactic
medium, so that any GCs formed after baryonic structure formation
resumes will be comparatively metal-rich.

The slope of the expected $T_{\rm blue}$ trend depends on the redshift
at which the first epoch of GC formation ended. Three curves shown in
Fig.~\ref{fig:tblue} illustrate this.  The curves come from an
extended Press-Schechter calculation \citep{ps74,lc93} done by G.\
Bryan (private communication). This type of calculation can be used to
determine the fraction of mass that is in collapsed halos of a given
mass at some early redshift, and that will later end up inside a more
massive halo today, at $z$ $=$ 0.  The specific calculation used to
create the curves in Fig.~\ref{fig:tblue} assumes that: galaxies are
formed via gravitational collapse and assembly of smaller halos that
collide and merge together over time; GCs can form within any one of
these smaller halos, as long as the halos have masses of at least
10$^8$~M$_{\odot}$; the number of metal-poor GCs that forms is
directly proportional to the fraction of a galaxy's mass that has
collapsed by a given redshift; and half the baryons within a given
galaxy halo will end up in the form of stars.  A constant
mass-to-light ratio was used to convert the final total mass of each
halo to a stellar mass, for the figure.  A $\Lambda$CDM cosmology was
assumed, with $\Omega_{m}$ = 0.3, $\Omega_{\Lambda}$ = 0.7, $\Omega_b
h^2$ $=$ 0.02, $h$ $=$ 0.65, and $\sigma_8$ $=$ 0.9.  Given these
assumptions, the relative number of metal-poor GCs within a given
galaxy at $z$ $=$ 0 depends on the redshift at which the metal-poor
GCs ceased to form.  Fig.~\ref{fig:tblue} shows the predicted $T_{\rm
blue}$ trend for three different truncation redshifts: $z$ $=$ 7, 11,
and 15.  The observed trend seems to fall (very roughly) between the
predicted trends for $z_{form} > 11$ and $z_{form} > 15$.  Much more
extensive simulations are needed in order to make rigorous predictions
for the relative numbers of metal-poor GCs --- and how these numbers
depend on the redshift of GC formation --- in galaxies over a range of
masses, environments, and merger/accretion histories.

Although the curves shown in the figure come from a fairly
straightforward calculation with several simplifying assumptions, they
show that in principle, a trend in $T_{\rm blue}$ such as we have
observed is generally consistent with a biased, hierarchical galaxy
formation scenario combined with the idea that the first generation of
GCs forms within a finite period in the early history of the Universe.
We should note here two factors that may influence the apparent
relationship between $T_{\rm blue}$ and galaxy stellar mass; these are
discussed in more detail in RZ04 and \citet{rzs05}.  We used the
constant $M/L_V$ value from \citet{za93} to calculate the stellar mass
$M_G$ for the elliptical galaxies. (For the spiral galaxies, $M/L_V$
in \citet{za93} changes with morphological type.)  In actuality,
$M/L_V$ may have a luminosity dependence as steep as $L^{0.10}$ (Zepf
\& Silk 1996 and references therein).  As we concluded in RZ04 and
\citet{rzs05}, such a dependence is not sufficient to explain the
observed $T_{\rm blue}$ trend.  Destruction of GCs through dynamical
effects (e.g., dynamical friction, evaporation, tidal shocks) may also
affect the observed $T_{\rm blue}$-galaxy mass relation.  For example,
\citet{vesp00} find that destruction may be more efficient in
lower-mass galaxies, although this is dependent on the details of how
galaxy potentials vary as a function of galaxy mass (e.g., Fall \&
Zhang 2001).  We note finally that relatively few moderate- and
high-luminosity galaxies are included in Fig.~\ref{fig:tblue} and that
filling in the sparsely-populated regions of the figure would be
useful for helping to determine the amount of ``biasing'' that may be
reflected in today's metal-poor GC populations.  We plan to continue
our work on measuring global properties and total numbers of GC
systems of galaxies with a range of luminosities, morphologies, and
environments; with $T_{\rm blue}$ values for dozens of galaxies on a
figure like Fig.~\ref{fig:tblue}, we can more strongly constrain the
redshift of formation of the first generation of GCs and their host
galaxies.

\section{Summary}
\label{section:summary}

As part of a larger survey that uses wide-field CCD imaging to study
the GC systems of giant galaxies, we have acquired and analyzed WIYN
$BVR$ imaging data of five nearly edge-on spiral galaxies: NGC~2683,
NGC~3044, NGC~3556, NGC~4157, and NGC~7331.  Our results are as
follows:

1.  We unequivocally detect the GC systems of all the galaxy targets
except the Sc galaxy NGC~3044.  Given the magnitude depth of our
images, our inability to detect NGC~3044's GC system may suggest that
the galaxy has a low $S_N$, or that it lies beyond the 23~Mpc distance
estimated from its recession velocity.

2. We observed the GC systems of the target galaxies to projected
   radial distances of $\sim$6$-$9 arc minutes (corresponding to
   20$-$40~kpc, depending on the distance to the galaxy) from the
   galaxy centers.  The GC surface density in our derived radial
   distributions vanishes before the last data point, suggesting that
   we have observed the full radial extent of the galaxies' GC
   systems.  

3.  The projected radial extents of the GC systems of the target
    spiral galaxies range from $\sim$10$-$20~kpc.  Combining the
    current data set with measurements for five other spiral,
    elliptical, and S0 galaxies from the survey, we derive a coarse
    relationship between host galaxy mass and radial extent of the GC
    system. Such a relationship is valuable for planning observations
    in which the aim is to observe all or most of the spatial extent
    of a galaxy's GC system.

4. The estimated total numbers of GCs in the spiral galaxies analyzed
   for this survey range from $\sim$80$-$290; the mean $N_{GC}$ is
   170$\pm$40.  One of the galaxies presented here, NGC~2683, had a
   previously-published $S_N$ value that is 2.5 times larger than our
   measured value.  The weighted mean $S_N$ and $T$ values for the
   five spiral galaxies in the survey are, respectively, 0.8$\pm$0.2
   and 1.4$\pm$0.3.  These values are consistent with the
   corresponding values for the Milky Way and M31, which suggests that
   the spiral galaxies with the most thoroughly studied GC systems are
   representative of the GC systems of other giant spiral galaxies
   with similar masses, at least in terms of their relative numbers of
   GCs.

5.  We estimate the galaxy-mass-normalized specific frequency of blue
    (metal-poor) GCs ($T_{\rm blue}$) in each galaxy and then combine
    these results with other data from the survey and the literature.
    The data confirm our initial conclusion (based on fewer points)
    that the metal-poor GC populations in luminous ellipticals are too
    large to have formed via the straightforward merger of two or more
    spiral galaxies and their associated metal-poor GC
    populations. The data likewise confirm that $T_{\rm blue}$
    generally increases with host galaxy mass. By comparing the
    $T_{\rm blue}$ vs.\ galaxy mass data to results from a simple
    model, we show that the observed trend is generally consistent
    with the idea that the first generation of GCs formed in galaxies
    over a finite period, prior to some truncation redshift.

\begin{figure}
\plotone{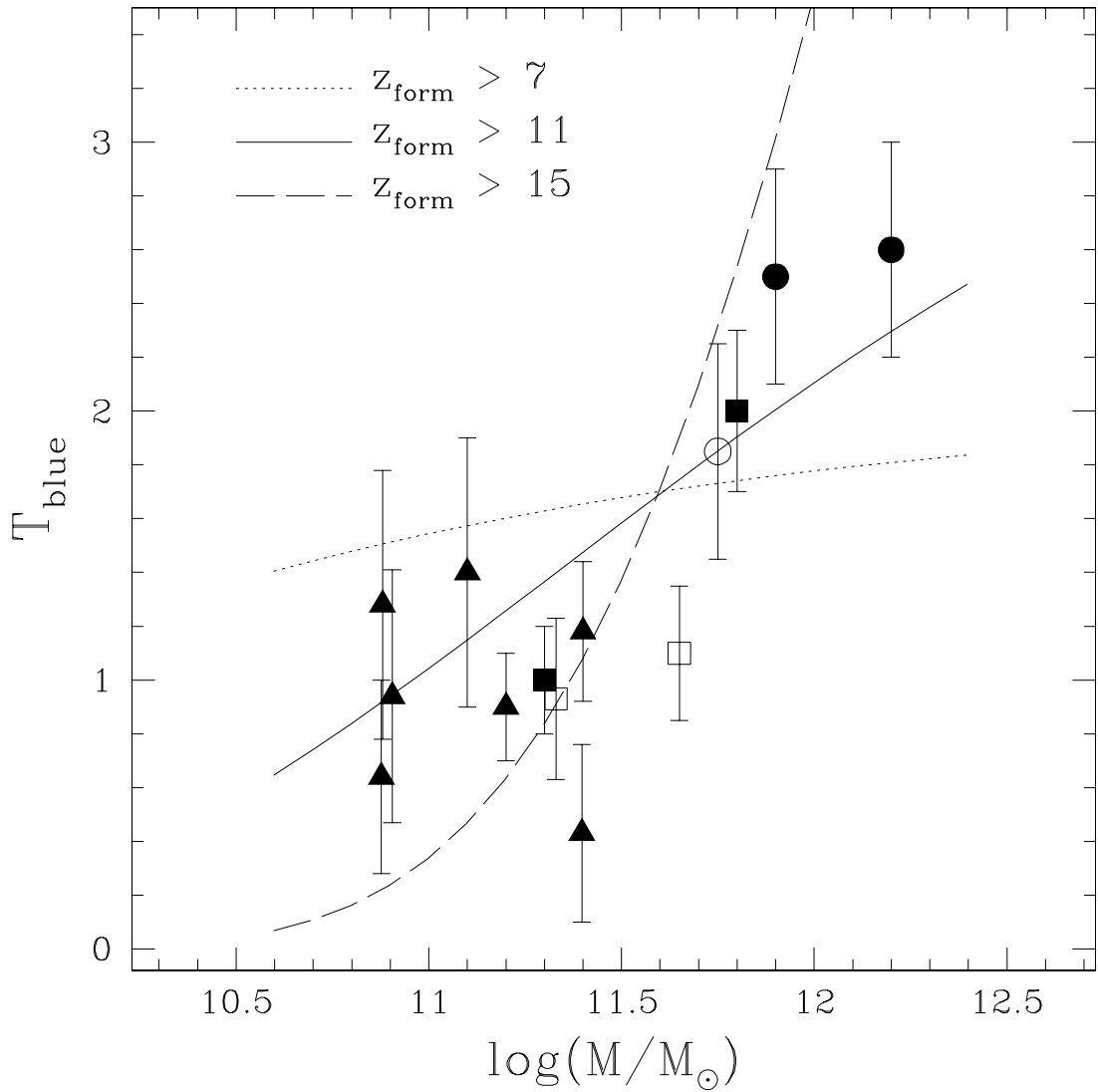}
\caption{\normalsize
Number of blue GCs normalized by host galaxy stellar mass
vs.\ the log of the galaxy stellar mass for 14 galaxies.  Circles are
cluster elliptical galaxies, squares are E/S0 field galaxies, and
triangles are field spiral galaxies.  Filled symbols denote the nine
galaxies from our wide-field survey, as well as the Milky Way and M31.
Open symbols are galaxies from other studies.  The curves (courtesy of
G.\ Bryan) show the expected trend if metal-poor GCs are formed prior
to $z$ of 7, 11, or 15, and assuming that the number of GCs is
proportional to the fraction of the galaxy mass that has assembled
prior to that redshift.}
\label{fig:tblue}
\end{figure}

\acknowledgments The research described in this paper was supported by
an NSF Astronomy and Astrophysics Postdoctoral Fellowship (award
AST-0302095) to KLR, NSF award AST 04-06891 to SEZ, and NASA Long Term
Space Astrophysics grant NAG5-12975 to AK.  We are grateful to the
Wesleyan University Astronomy Department for funding ANL while he
analyzed the data for NGC~7331.  We thank Greg Bryan for illuminating
discussions and for providing the model calculations shown in
Fig.~\ref{fig:tblue}.  We thank the staff at the WIYN Observatory and
Kitt Peak National Observatory for their assistance at the telescope.
We also thank Enzo Branchini, who provided some estimated distances
for the target galaxies based on a model of the local velocity flow.
Finally, we thank the anonymous referee for valuable comments and
suggestions that improved the quality of the paper.  This research has
made use of the NASA/IPAC Extragalactic Database (NED) which is
operated by the Jet Populsion Laboratory, California Institute of
Technology, under contract with the National Aeronautics and Space
Administration.



%
%


\clearpage

\clearpage
\tabletypesize{\normalsize}
\begin{deluxetable}{lccrr}
\tablecaption{Basic Properties of the Target Galaxies}
\tablewidth{260pt}
\tablehead{\colhead{Name}&\colhead{Type}&\colhead{$m-M$} &\colhead{Distance} &\colhead{$M_V^T$}\\
\colhead{} & \colhead{} & \colhead{} & \colhead{(Mpc)} & \colhead{}}
\startdata
NGC~2683 & Sb & 29.44 & 7.7 & $-$20.5 \\ 
NGC~3044 & Sc & 31.83 & 23.2 & $-$21.0 \\ 
NGC~3556  & Sc & 30.46  & 7.1 & $-$21.2 \\ 
NGC~4157 & Sb & 30.84 & 14.7 & $-$20.4 \\ 
NGC~7331 & Sb & 30.59 & 13.1 & $-$21.7 \\ 
\enddata
\tablecomments{Distances to NGC~2683, NGC~4157, and NGC~7331 are from
  Tonry et al.\ (2001) (surface brightness fluctuations).  Distances
  to NGC~3044 and NGC~3556 are from combining the recession velocity with
  respect to the cosmic microwave background from RC3 (deVaucouleurs
  et al.\ 1991) with H$_0$ $=$ 70 km~s$^{-1}$~Mpc$^{-1}$.
  Magnitudes are from combining $V_T^0$ from RC3 (deVaucouleurs et
  al.\ 1991) with $m-M$.}
\protect\label{table:galaxy properties}
\end{deluxetable}

\begin{deluxetable}{lccrrr}
\tablecaption{WIYN Observations of Target Galaxies}
\tablewidth{370pt}
\tablehead{\colhead{Galaxy} &\colhead{Date}&\colhead{Detector}
  &\multicolumn{3}{c}{Exposure Times (seconds)} \\
\colhead{} & \colhead{} & \colhead{} & \colhead{$B$} & \colhead{$V$} & \colhead{$R$}}
\startdata
NGC~2683 & Jan 2001 & Minimosaic & 3 x 2100 & 3 x 2000 & 3 x 1500 \\
NGC~3044 & Jan 2000 & Minimosaic & 3 x 1800 & 3 x 1800 & 3 x 1800 \\
NGC~3556  & Apr 2000 & Minimosaic &3 x 2100 & 4 x 2000 & 3 x 1500 \\
NGC~4157 & Jan 2000 & Minimosaic & 3 x 1800 & 3 x 1500 & 3 x 1500 \\
NGC~7331 & Oct 1999 & S2KB & 3 x 1800 & 3 x 1800 & 3 x 1500 \\
\enddata
\protect\label{table:wiyn observations}
\end{deluxetable}

\begin{deluxetable}{lrrrrr}
\tablecaption{Aperture Corrections Used for Photometry of Sources in
  WIYN Images}
\tablewidth{500pt}
\tablehead{\colhead{} & \colhead{NGC~2683}& \colhead{NGC~3044}& \colhead{NGC~3556}& \colhead{NGC~4157}& \colhead{NGC~7331}}
\startdata
$B......$ & $-$0.234 $\pm$ 0.002 & $-$0.290 $\pm$ 0.003 & $-$0.343 $\pm$ 0.009& $-$0.278 $\pm$ 0.009& $-$0.267 $\pm$ 0.003\\
$V......$ & $-$0.294 $\pm$ 0.004 & $-$0.224 $\pm$ 0.004 & $-$0.270 $\pm$ 0.006& $-$0.206 $\pm$ 0.007& $-$0.353 $\pm$ 0.003\\
$R......$ & $-$0.152 $\pm$ 0.003 & $-$0.124 $\pm$ 0.004 & $-$0.230 $\pm$ 0.005& $-$0.280 $\pm$ 0.007& $-$0.330 $\pm$ 0.002\\
\enddata
\protect\label{table:aper corr}
\end{deluxetable}

\begin{deluxetable}{lccccc}
\tablecaption{Extinction Corrections Used for Photometry of Sources in
WIYN Images}
\tablewidth{380pt}
\tablehead{\colhead{} & \colhead{NGC~2683}& \colhead{NGC~3044}& \colhead{NGC~3556}& \colhead{NGC~4157}& \colhead{NGC~7331}}
\startdata
$A_B......$ & 0.140 & 0.108 & 0.068& 0.091& 0.394\\
$A_V......$ & 0.108 & 0.084 & 0.055& 0.069& 0.303\\
$A_R......$ & 0.087 & 0.068 & 0.042& 0.057& 0.245\\
\enddata
\protect\label{table:ext corr}
\end{deluxetable}

\begin{deluxetable}{lrrrr}
\tablecaption{50\% Completeness Limits of WIYN Images}
\tablewidth{300pt}
\tablehead{\colhead{} & \colhead{NGC~2683}& \colhead{NGC~3556}& \colhead{NGC~4157}& \colhead{NGC~7331}}
\startdata
$B......$ & 25.2  & 25.0  & 24.5  & 24.8 \\
$V......$ & 24.8  & 24.7  & 24.3  & 24.6 \\
$R......$ & 24.6  & 24.7  & 23.6  & 24.1 \\
\enddata
\protect\label{table:completeness}
\end{deluxetable}

\begin{deluxetable}{llrcc}
\tablecaption{HST WFPC2 Observations Analyzed for this Study}
\tablewidth{350pt}
\tablehead{\colhead{Proposal ID} & \colhead{Target Name}&
  \colhead{Exp Time}&
\colhead{Ang Sep} & \colhead{Filter}\\
\colhead{} & \colhead{} & \colhead{(sec)} & \colhead{($\arcm$)} & \colhead{}}
\startdata
NGC 2683: & & & & \\
5446    &   NGC~2683 &	160	 & 0.7&	F606W\\
8199    &   NGC~2683 &  2600 & 1.9& F814W\\
\cr
\tableline					
\cr
NGC 3556: & & & & \\
5446   	&	NGC~3556& 160 	 & 0.4 &	F606W\\
\cr
\tableline					
\cr
NGC 7331: & & & & \\
5397/6294 & NGC~7331 & 40400 & 3.5 & F555W\\
5397/6294 & NGC~7331 & 9600 & 3.5 & F814W\\
8805      & Any  & 2500 & 5.2 & F606W\\
\enddata
\protect\label{table:hst data}
\end{deluxetable}

\begin{deluxetable}{lr}
\tablecaption{Corrected Radial Profile of the GC System of NGC~2683}
\tablewidth{160pt}
\tablehead{\colhead{Radius} & \colhead{Surface Density}\\
\colhead{(arcmin)} & \colhead{(arcmin$^{-2}$)}}
\startdata
1.2 & 5.90 $\pm$ 1.75\\
2.0 & 1.34 $\pm$ 0.66\\
3.0 & 0.80 $\pm$ 0.54\\
4.0 & $-$0.04 $\pm$ 0.35\\
5.0 & 0.18 $\pm$ 0.38\\
6.0 & $-$0.12 $\pm$ 0.32\\
7.0 & $-$0.02 $\pm$ 0.33\\
7.9 & $-$0.03 $\pm$ 0.36\\
8.9 & 0.02 $\pm$ 0.46\\
\enddata
\label{table:profile n2683}
\end{deluxetable}

\begin{deluxetable}{lr}
\tablecaption{Corrected Radial Profile of the GC System of NGC~3556}
\tablewidth{160pt}
\tablehead{\colhead{Radius} & \colhead{Surface Density}\\
\colhead{(arcmin)} & \colhead{(arcmin$^{-2}$)}}
\startdata
1.6 & 7.78 $\pm$ 2.21\\
2.6 & 3.77 $\pm$ 1.02\\
3.5 & 1.66 $\pm$ 0.56\\
4.5 & 0.75 $\pm$ 0.35\\
5.5 & $-$0.02 $\pm$ 0.17\\
6.5 & 0.00 $\pm$ 0.18\\
7.5 & 0.01 $\pm$ 0.19\\
8.5 & 0.01 $\pm$ 0.19\\
\enddata
\label{table:profile n3556}
\end{deluxetable}

\begin{deluxetable}{ll}
\tablecaption{Corrected Radial Profile of the GC System of NGC~4157}
\tablewidth{200pt}
\tablehead{\colhead{Radius} & \colhead{Surface Density}\\
\colhead{(arcmin)} & \colhead{(arcmin$^{-2}$)}}
\startdata
1.1 & 1.88 $\pm$ 1.12\\
2.1 & 1.00 $\pm$ 0.61\\
3.0 & 0.56 $\pm$ 0.42\\
4.0 & 0.35 $\pm$ 0.33\\
5.0 & 0.09 $\pm$ 0.27\\
6.0 & 0.00 $\pm$ 0.26\\
7.0 & 0.39 $\pm$ 0.38 (0.20 $\pm$ 0.33)\\
8.0 & 0.98 $\pm$ 0.51 ($-$0.16 $\pm$ 0.21)\\
8.9 & 0.17 $\pm$ 0.38\\
\enddata
\tablecomments{Values in parentheses are the surface density when the
  seven loosely-grouped GC candidates located in the galaxy's outer halo is
  removed.}
\label{table:profile n4157}
\end{deluxetable}

\begin{deluxetable}{lr}
\tablecaption{Corrected Radial Profile of the GC System of NGC~7331}
\tablewidth{200pt}
\tablehead{\colhead{Radius} & \colhead{Surface Density}\\
\colhead{(arcmin)} & \colhead{(arcmin$^{-2}$)}}
\startdata
1.9 &  3.30 $\pm$ 1.57 \\
2.8 & 0.75 0.$\pm$ 89 \\
3.8 & 1.88 $\pm$ 0.96 \\
4.8 & $-$0.12 $\pm$ 0.60 \\
5.7 & 0.19 $\pm$ 0.73 \\
\enddata
\label{table:profile n7331}
\end{deluxetable}

\begin{deluxetable}{lcccc}
\tablecaption{Coefficients from Fitting Radial Profile Data}
\tablewidth{370pt}
\tablehead{
\colhead{} & \multicolumn{2}{c}{deVaucouleurs Law} &
\multicolumn{2}{c}{Power Law}\\
\cline{2-3} \cline{4-5}\\
\colhead{Galaxy} & \colhead{a0}& \colhead{a1} & \colhead{a0}& \colhead{a1}}
\startdata
NGC~2683 & 4.28 $\pm$ 0.77 & $-$3.39 $\pm$ 0.68 & 0.93 $\pm$ 0.13 & $-$2.35 $\pm$ 0.47\\
NGC~3556 & 4.48 $\pm$ 0.66 & $-$3.13 $\pm$ 0.52 & 1.43 $\pm$ 0.17 & $-$2.27 $\pm$ 0.38\\
NGC~4157 \tablenotemark{\dag} & 0.74 $\pm$ 0.50 & $-$0.62 $\pm$ 0.35 & 0.16 $\pm$ 0.17 & $-$0.51 $\pm$ 0.27\\
         & 2.02 $\pm$ 0.67 & $-$1.71 $\pm$ 0.53 & 0.34 $\pm$ 0.18 & $-$1.27 $\pm$ 0.39\\
NGC~7331 & 2.33 $\pm$ 1.28 & $-$1.57 $\pm$ 0.99 & 0.81 $\pm$ 0.34 & $-$1.16 $\pm$ 0.74\\
\enddata
\tablenotetext{\dag}{The first set of coefficients for NGC~4157 is
  for fits to the radial profile data including a group of seven GC
  candidates in the galaxy's halo; the second set is for fits to 
  the profile with those seven candidates removed.}
\protect\label{table:profile coefficients}
\end{deluxetable}

\begin{deluxetable}{lccc}
\tablecaption{Radial Extents of GC Systems of Survey Galaxies}
\tablewidth{330pt}
\tablehead{\colhead{Galaxy} &\colhead{Mass} 
&\colhead{GC System Extent} &\colhead{Reference}\\
 \colhead{} & \colhead{[log(M/M$_\odot$)]} & \colhead{(kpc)} & \colhead{}}
\startdata
NGC~2683 & 10.9 & 9$\pm$3 & 4\\
NGC~3379 & 11.3 & 34$\pm$4 & 3\\
NGC~3556 & 11.1 & 20$\pm$4 & 4\\
NGC~4157 & 10.9 & 21$\pm$4 & 4\\
NGC~4406 & 11.9 & 83$\pm$6 & 3\\
NGC~4472 & 12.2 & 102$\pm$7 & 1\\
NGC~4594 & 11.8 & 54$\pm$5 & 3\\
NGC~7331 & 11.4 & 18$\pm$4 & 4\\
NGC~7814 & 10.9 & 13$\pm$4 & 2\\
\enddata
\tablerefs{(1) Rhode \& Zepf 2001; (2)  Rhode
  \& Zepf 2003; (3) Rhode \& Zepf 2004; (4) this paper.}
\protect\label{table:extents}
\end{deluxetable}

\begin{deluxetable}{lrccc}
\tablecaption{Total Numbers and Specific Frequencies of GC Systems}
\tablewidth{290pt}
\tablehead{\colhead{Galaxy}
&\colhead{$N_{GC}$}&\colhead{$S_N$} 
&\colhead{$T$}&\colhead{$T_{\rm blue}$}\\
\colhead{} & \colhead{} & \colhead{} & \colhead{} & \colhead{}}
\startdata
NGC~2683 & 120$\pm$40 & 0.8$\pm$0.4 & 1.4$\pm$0.7  & 0.9$\pm$0.5 \\
%
NGC~3556 & 290$\pm$80 & 0.9$\pm$0.4 & 2.2$\pm$0.9 & 1.4$\pm$0.5 \\
NGC~4157 & 80$\pm$20 & 0.6$\pm$0.3 & 1.1$\pm$0.6 & 0.6$\pm$0.4 \\
NGC~7331 & 210$\pm$130 & 0.5$\pm$0.4 & 0.9$\pm$0.7 & 0.4$\pm$0.3 \\
\enddata
\protect\label{table:total numbers}
\end{deluxetable}

\end{document}